\pdfoutput=1
\documentclass{sig-alternate} 
\usepackage{mathptmx} 

\newcommand{\ignore}[1]{}
\usepackage{fancyhdr}

\usepackage[sort,nocompress]{cite}
\usepackage[final]{microtype}
\usepackage[normalem]{ulem}
\usepackage[hyphens]{url}
\usepackage{tcolorbox}

\usepackage{fancyhdr,graphicx,amsmath,amssymb}
\usepackage[ruled,vlined]{algorithm2e}

\usepackage[bookmarks=true,breaklinks=true,letterpaper=true,colorlinks,linkcolor=black,citecolor=blue,urlcolor=black]{hyperref}

\newcommand{\microsubmissionnumber}{339}
\ignore{
\fancypagestyle{firstpage}{
  \fancyhf{}
  
  \fancyhead[C]{\vspace{15pt}\normalsize{MICRO 2019 Submission
      \textbf{\#\microsubmissionnumber} -- Confidential Draft -- Do NOT Distribute!!}} 
  \fancyfoot[C]{\thepage}
}
}
\title{\vspace{-0.45 in}Lookout for Zombies: Mitigating Flush+Reload Attack \\ on Shared Caches by Monitoring Invalidated Lines\vspace{-0.5in}}

\author{
Gururaj Saileshwar and Moinuddin K. Qureshi\vspace{0.1in}\\
Georgia Institute of Technology\vspace{0.1in}\\
\textit{\large{\{gururaj.s,moin\}@gatech.edu}}\vspace{0.1in}
}

\begin{document}
\abovedisplayskip=0.7cm
\abovedisplayshortskip=-0.1cm
\belowdisplayskip=0.7cm
\belowdisplayshortskip=0.4cm
 
\maketitle

\begin{abstract}


OS-based page sharing is a commonly used optimization in modern systems to reduce memory footprint. Unfortunately, such sharing can cause Flush+Reload cache attacks, whereby a spy periodically flushes a cache line of shared data (using the \texttt{clflush} instruction) and reloads it to infer the access patterns of a victim application. Current proposals to mitigate Flush+Reload attacks are impractical as they either disable page sharing, or require application rewrite, or require OS support, or incur ISA changes.  Ideally, we want to tolerate attacks without requiring any OS or ISA support and while incurring negligible performance and storage overheads.





This paper makes the key observation that when a cache line is invalidated due to a {\em Flush-Caused Invalidation (FCI)}, the tag and data of the invalidated line are still resident in the cache and can be used for detecting Flush-based attacks. We call lines invalidated due to FCI as {\em Zombie} lines. Our design explicitly marks such lines as Zombies, preserves the Zombie lines in the cache, and uses the hits and misses to Zombie lines to tolerate the attacks. We propose {\em Zombie-Based Mitigation (ZBM)}, a simple hardware-based design that successfully guards against attacks by simply treating hits on Zombie-lines as misses to avoid any timing leaks to the attacker.  We analyze the robustness of ZBM using three spy programs: attacking AES T-Tables, attacking RSA Square-and-Multiply, and Function Watcher (FW), and show that ZBM successfully defends against these attacks. Our solution requires negligible storage (4-bits per cacheline), retains OS-based page sharing, requires no OS/ISA changes, and does not incur slowdown for benign applications.\ignore{ We also discuss how our solution can also tolerate other flush-based attacks on cache and memory.}

\vspace{0.1in}



\end{abstract}

\ignore{
FLOW:

Data sharing needed but leads to cache attacks

Prime+Probe eviction based attacks -- can be tolerated

Flush-based attacks directly target shared data and harder to tolerate.
Show figure.

The existing solutions are messy.  Insight.

Zombie, mark zombie, protect zombie, install zombie, hit on zombie. Tolerates in hardware

We can do ZBD (detection in hardware) and mitigation in OS as well. 

While we use ZBM only for Flush-Reload attack, it can also be extended to other attacks such as Flush+Flush attacks and Coherence-Based attacks.  We discuss this in Section XYZ.

}

\section{Introduction}
\label{sec:intro}
\vspace{0.15 in}
Caches alleviate the long latency of main memories by providing data with low latency.  Unfortunately, the timing difference between the cache hit and a cache miss can be used as a {\em side-channel} by an adversary to infer the access pattern and obtain unauthorized information from the system.  Such cache attacks have been demonstrated to leak sensitive information like encryption keys~\cite{Bernstein} and user browsing activity in the cloud~\cite{Hornby2016}. More recent attacks like Spectre~\cite{spectre} and Meltdown~\cite{meltdown} use  cache-based side-channels to convert sensitive data accessed illegally into discernible information. Given their widespread impact, there is a pressing need for efficient mitigation of such cache side-channels. At the same time, it is also important to preserve the benefits of the performance optimizations, that are often the cause of these side channels.

\ignore{
Caches alleviate the long latency of main memories by providing data with low latency.  Unfortunately, the timing difference between the cache hit and a cache miss can be used as a {\em side-channel} by an adversary to infer the access pattern and obtain unauthorized information from the system. The recently disclosed Spectre~\cite{spectre}, Meltdown~\cite{meltdown}, and ForeShadow~\cite{foreshadow} vulnerabilities use such cache-based side channels to convert the unauthorized data value into discernible information.  While cache attacks have been demonstrated in the past at a smaller scale, the widespread impact of the recent vulnerabilities highlights the pressing need for efficient mitigation of cache side-channels. At the same time, it is also important to preserve the benefits of the performance optimizations, that are often the cause of these side channels.
}
\ignore{
Caches alleviate the long latency of main memories by providing data with low latency.  Unfortunately, the timing difference between the cache-hit and a cache-miss can be used as a {\em side-channel} by an adversary to infer the access pattern and obtain unauthorized information from the system. The recently disclosed Spectre~\cite{spectre}, Meltdown~\cite{meltdown}, and ForeShadow~\cite{foreshadow} vulnerabilities rely on such cache-based side channels to convert the unauthorized data value into discernible information.  While cache attacks have been demonstrated in the past at a smaller scale, the recent vulnerabilities show that cache attacks can affect hundreds of millions of processor systems, and highlight the need to develop efficient solutions to mitigate cache attacks. These attacks also highlight the need to rethink performance optimizations for potential security vulnerabilities. 
}

OS-based page sharing~\cite{KSM,MemDedupVmware} is a commonly used optimization in modern systems to avoid redundant copies of pages across applications (such as for library code) or for supporting multiple copies of the same data pages (by deduplicating data pages). Such sharing allows different programs accessing the same library code to get routed to the same OS page. While OS-based page sharing is useful for memory capacity, such sharing leads to cache side channels between processes, even though the pages are shared only in read-only mode.

\ignore{
OS-based page sharing is a common optimization used in modern systems to avoid redundant copies of pages across applications (such as for library code) or for having multiple copies of the same data pages (by de-duplication of data pages). Such sharing allows different programs accessing the same library code to get routed to the same OS page. While OS-based page sharing is beneficial for memory capacity, such sharing leads to cache side channels between applications, even if the pages are shared only in read-only mode.

}


In this paper, we study cross-core Flush+Reload attack~\cite{yaromFlushReload}, a highly effective attack that uses the OS-shared pages between the attacker and the victim.  In this attack, the attacker periodically evicts a cache line that is shared with the victim from the cache using the \texttt{clflush} instruction, waits for a predefined interval, and then uses a timing check to infer whether the victim application accessed that line in the interim. 

\begin{figure*}[htb] 
  	\centering
		\includegraphics[width=6.95in]{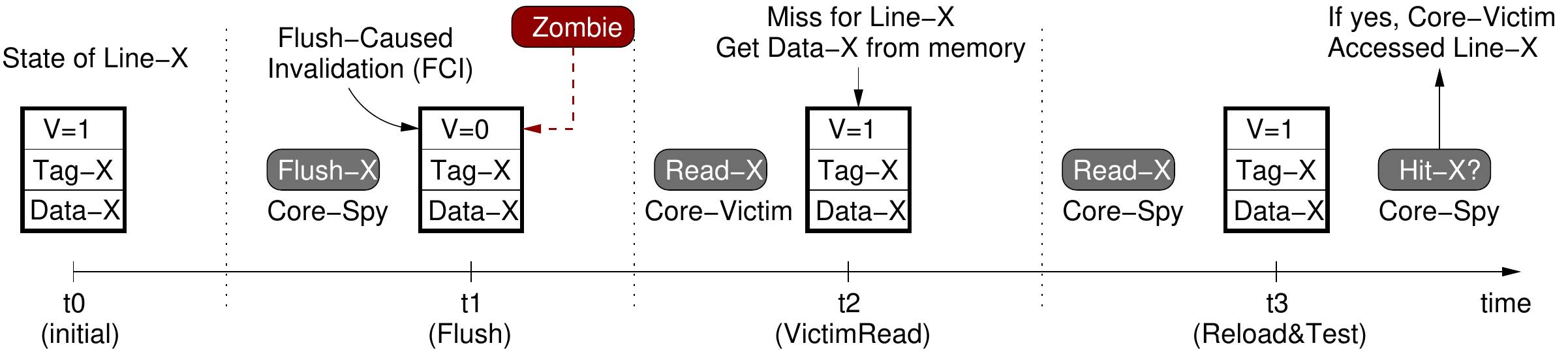}
      \vspace{-0.15 in}
	\caption{Example of Flush+Reload attack on a shared line.  The spy program on Core-Spy flushes Line-X from the cache and waits. The victim program on Core-Victim accesses Line-X at t2. At t3, the spy accesses Line-X and uses a timing test to check for a cache hit, and infers that the victim accessed Line-X. Our solution marks lines invalidated due to Flush-Caused Invalidation (FCI) as a "Zombie" line and monitors hits and misses to Zombie lines to tolerate attacks.}
   \vspace{-0.1 in}
	\label{fig:intro} 
\end{figure*}

We explain Flush+Reload attack with an example.  Figure~\ref{fig:intro} shows an example of Flush+Reload attack for a given line, Line-X, of shared data. At time t0, the state of Line-X is shown (V denoting the valid bit, the tag Tag-X and the data Data-X). Line-X is shared between the spy and the victim. At time t1, the spy program (running on Core-Spy) flushes this line using a \texttt{clflush} instruction. This instruction invokes a {\em Flush-Caused Invalidation (FCI)} that invalidates the given line in the cache.  The spy program then waits for some time.  At a later time t2, the victim program (running on Core-Victim) may access the Line-X.  This access will miss in the cache and the cache controller will retrieve the data for Line-X (Data-X) from memory and install it in the cache. At time t3, the Core-Spy accesses Line-X and measures the time for the access to determine if Core-Victim accessed the line during the waiting time. The spy can use the patterns of hits and misses to infer secret information of the victim.


\ignore{

Eviction based cache attacks can be classified into two categories: conflict-based attacks and flush-based attacks.  Conflict-based attacks (such as Prime+Probe~\cite{PrimeProbe}) cause eviction of the given line by accessing lines that map to the same set. Fortunately, caches can be protected against conflict-based attacks using cache-space preservation (reserving some portion of the cache exclusively for the given core)~or randomizing the location of the line in the cache~\cite{RPCache,NewCache,CEASER}. 
}

Prior solutions\cite{RPCache,NoMo,kongZhou,Page:SideChannel,NewCache,CEASER} that are highly effective at guarding the cache against conflict-based attacks  (such as Prime+Probe\cite{PrimeProbe}) become ineffective for Flush-based attacks, as the attacker can precisely invalidate a given line from the cache, without relying on any conflicts. Current proposals to mitigate Flush-based cache attacks primarily rely on OS-based or software-based techniques. For example, cache attacks on shared data can be avoided by disabling sharing of critical pages or by dynamically replicating shared pages~\cite{CacheBar}, or rewriting the application using transaction memory semantics~\cite{tmem:usenix2017} or using performance monitoring counters to detect deviation in behavior of the application~\cite{PerfCtr1,PerfCtr2}.  Ideally, we want to protect Flush-based attacks while retaining page sharing, without requiring a rewrite of victim applications, and without relying on profile information for the performance monitoring counters. 

Architectural solutions for mitigating Flush-based cache attacks fall into two categories: {\em Restriction} and {\em Duplication}. SHARP~\cite{SHARP} proposes to restrict the use of {\texttt{clflush}} in user-mode on read-only pages. However, such a solution requires changing the ISA definition of {\texttt{clflush}} and is not backwards compatible. NewCache~\cite{Ruby:MICRO2017} proposes to access the cache line with Process-ID and the line-address -- this creates duplicate copies of the line in the NewCache if two different processes concurrently access the same line, thereby avoiding the flush of one process from evicting the line of another. Unfortunately, such line duplication is incompatible with inter-process communication. DAWG~\cite{DAWG} also proposes to create a replica of the shared line for each Domain-ID, and the communicating threads (or processes) must be grouped within the same Domain. However, DAWG requires OS support for cache management and for grouping applications into security domains. The goal of this paper is to develop  a hardware-based solution that does not require any OS support, does not require ISA changes, does not restrict inter-process communication, and still mitigates the attack while incurring negligible overheads.  Our solution leverages the hardware events that are inherent in such attacks. 

This paper makes the key observation that when a cache line is invalidated due to {\em Flush-Caused Invalidation (FCI)}, the tag and data of the invalidated line are still resident in the cache and can be used for detecting Flush-based attacks. For example, in Figure~\ref{fig:intro} at time t1, the FCI resets the valid bit of Line-X while still leaving Tag-X and Data-X in the cache line.   We call such lines invalidated due to FCI that still contain meaningful Tag and Data information as {\em Zombie} lines. 

Our solution uses Zombie lines to detect Flush-based attacks and has four parts. (1) \textit{Mark the Zombies}: To enable monitoring of Zombie lines, we extend the tag entry of the cache line to include a bit (Z-bit), to identify it as a Zombie line (at time t2, Line-X would have Z=1). (2) \textit{Protect the Zombies}: Conventional replacement algorithms are designed to preferentially pick invalid lines on a cache miss. To protect the Zombie lines, we modify the replacement policy to treat an invalid Zombie line similar to a valid line, and evict the line only when it would have been naturally evicted if it had not received an FCI. (3) \textit{Deterministic Victim Selection on Zombie Miss}: On a cache miss, if there is a tag hit on the invalid Zombie line, we know that the line was recently invalidated due to an FCI. We detect such cases, and modify the replacement algorithm to always pick the invalid-zombie line as the victim on line install in such scenarios. We retain Z=1 if the data obtained from memory was identical to the data resident in the line, as this denotes a case where the flush of the line was done unnecessarily. (4) \textit{Mitigate on Zombie Hit}: A hit to a valid zombie line invokes mitigating actions that tolerate the attack by avoiding leakage of timing information.  

We propose {\em Zombie-Based Mitigation (ZBM)}, a simple hardware-solution to tolerate cross-core Flush+Reload attacks. ZBM simply treats Zombie hits as cache misses, that incur the latency of a memory access -- by inserting a dummy memory access, waiting for it to complete, and only then returning the data to the requestor. This eliminates any timing channel for the attacker as both cache hits and misses incur the same latency once the line is marked as a zombie line. 

We analyze robustness of ZBM using three spy programs (similar to prior works~\cite{SHARP,tmem:usenix2017}):\ignore{that are based on cross-core Flush+Reload attacks:} (1)~Attacking AES T-Tables (2)~Attacking Square-and-Multiply algorithm of RSA, and (3)~Monitoring a victim's function usage with {\em Function Watcher}.  We demonstrate that ZBM  successfully mitigates these attacks by closing the timing channel of cache line flushes. 

ZBM only requires 1-bit per cache line (Z-bit) in the shared L3 cache (private L1/L2 caches are unchanged) and causes no slowdown for applications without flush and reload of identical contents. For niche applications like persistent memory, non-coherent I/O, etc. where frequent flushes are possible,\ignore{ we analytically model the worst-case slowdown with ZBM. We then extend our design to \textit{ZBMx} to avoid slowdown for such applications with repeated flushes, requiring 3-bits per cache line, to track the flush issuing core, in addition to the Z-bit.} we avoid slowdown by extending our design to \textit{ZBMx} that also tracks the flush causing core-id (3 bits per cache line) in addition to the Z-bit per line. With ZBM and ZBMx, we are able to avoid slowdown for all typical non-attack scenarios.  

\ignore{Moreover, ZBM allows unrestricted OS-based page sharing and requires no OS/ISA changes.}

\vspace{0.075 in}

Overall this paper makes the following contributions:

\begin{enumerate}
    \item To the best of our knowledge, this is the first paper to use the inherent state of the cache to detect Flush-based cache attacks. We mark the flushed lines as zombies and check for reloading of identical content on install.
    
    \item We propose a simple hardware mitigation of cross-core Flush+Reload attack by servicing hits to zombie lines as misses. Our solution mitigates the attack and incurs no slowdown for typical benign applications.\ignore{ that do not flush and then reload identical contents. }
    
    \item We show our solutions (ZBM and ZBMx) can be implemented with negligible storage (1-4 bits per cache line), and without any changes to the OS, software, or ISA.
    
    \ignore{
    
    \item We show how our insight of tracking invalidated lines can be extended to tolerate other flush-based attacks, such as Flush+Flush attack, Coherence-Based attack, Row-Hammer attack, and DRAM Row-Buffer attack.
    }

\end{enumerate}

\ignore{
While ZBM is a hardware-based solution, we also discuss how tracking of Zombie lines can enable a hybrid approach, where the hardware detects the attack and OS mitigates the attack. {\em Zombie-Based Detection (ZBD)} of attacks is performed by tracking the core that performs the flush and the core that reloads the zombie lines to precisely identify the spy and victim programs. Once a given number of flush and reloads, ZBD informs the OS of the potential attack and the cores corresponding to spy and victim processes. The OS can  isolate the victim and spy into two systems with independent memory domains (such as different machines in a cloud environment) or into two different time slices on the same system.  

We discuss implications of our solution on Prime+Probe attacks and Denial of Service attacks and show that our solutions do not worsen these attacks.  We also discuss how the Zombie-based detection can be extended to detect (and possibly mitigate) other cache attacks, like Flush+Flush attack and Coherence-Based attacks and even memory timing attacks.  
}

\newpage

\ignore{
FLOW
- Attack Setup and RSA (Square Multiply Algorithm)
- Eviction Based Attacks and Mitigation
- Flush Based Cache Attacks and Mitigation
- Goal
}

\section{Background and Motivation}

Modern computing systems rely on sharing the resources such as the last-level cache and main memory across processes to improve efficiency.  While such sharing is useful for performance and reducing cost, the sharing of resources can create side channels.  We discuss the background on OS-based page sharing, the settings in which such sharing could lead to a cross-core attack, and typical forms of cache attack.

\subsection{OS-Based Page Sharing}
OS-based page sharing reduces memory footprint by removing redundant copies of identical pages.  Such sharing is essential for sharing of the text segment of executable files between processes and for using shared libraries.  Furthermore, {\em memory deduplication}  is a popular technique to explicitly identify memory pages containing identical contents in unrelated pages and coalesce these pages into a single unit. Memory deduplication has been implemented in a variety of systems\cite{KSM,MemDedupVmware}, including VMWare and PowerVM hypervisors, and in Linux and Windows. While read operations are permitted on deduplicated pages, a write operation results in a copy-on-write exception (to replicate the page and map it into the process).  While page sharing is useful for effectively utilizing memory capacity and for enabling shared libraries, it can lead to side channels which can be exploited by an attacker, even though the pages are shared in read-only mode.

\subsection{Attack Model}

In this paper, we focus on cross-core cache attacks, where the victim  and the spy are executing on separate cores of a multi-core processor.  This is a safe assumption in cloud computing environment, where the non-trusting applications are not concurrently scheduled on the same core.  

\begin{figure}[htb] 
  	\centering
  	\vspace{0.25 in}
	\includegraphics[width=3.50in]{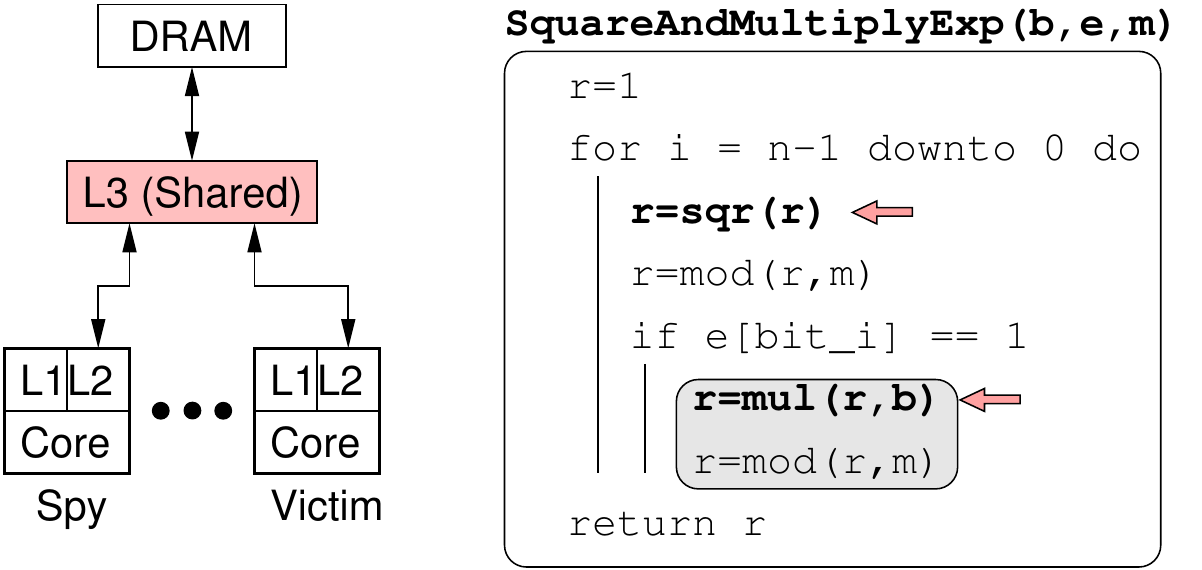}
	\caption{Setup for Cross-Core Attacks (left) and Exponentiation  based on Square and Multiply (right)}
	\label{fig:setup} 
\end{figure}

Figure~\ref{fig:setup} shows our system configuration, which contains a multi-core processor with private L1 and L2 caches, and a shared L3 cache.  The L3 cache is inclusive and evictions from the L3 cache cause evictions from L1 and L2 (if the line is present).  As L3 cache is the point at which resources are shared, the adversary tries to orchestrate evictions in the L3 cache and monitor the hits in the L3 cache to observe the behavior of the victim program. Such cache attacks have been used to infer secrets such as the keys for AES~\cite{Bernstein}.

\subsection{Example: RSA Square-And-Multiply}
\label{sec:rsa_bg}
Figure~\ref{fig:setup} shows the code for the square-and-multiply algorithm used in RSA implementation of GnuPG version 1.4.13 that is vulnerable to cache attacks (recent versions have moved towards secure implementations). The algorithm computes $(b^e)\text{\space mod\space} m$\ignore{"b raised to the power of e, followed by mod m"}, i.e. ``\textit{b} raised to the power of \textit{e}, modulo \textit{m}''. The algorithm iterates from the most-significant bit of \textit{e} (the secret key) to the least-significant bit, always performing a square operation (top arrow) and performing the multiply operation (bottom arrow) only if the bit is "1". By observing the access pattern for lines containing the square (\textit{sqr}) and multiply (\textit{mul}) functions, the spy can infer the bits of the secret (\textit{e})  -- \textit{sqr} followed by \textit{sqr} is a 0 whereas \textit{sqr} followed by \textit{mul} is a 1. The lines containing the instructions for the \textit{sqr} and the \textit{mul} functions are called {\em probe addresses}.

The spy can infer the access pattern of the victim by causing an eviction of the probe address, waiting, and then testing if the probe address was accessed by the victim (by checking if a hit is obtained for the probe address). Depending on how the evictions are performed, cache attacks can be classified as either (a) conflict-based attacks or (b) flush-based attacks.

\ignore{

\begin{figure}[htb] 
  	\centering
	\includegraphics[width=3in]{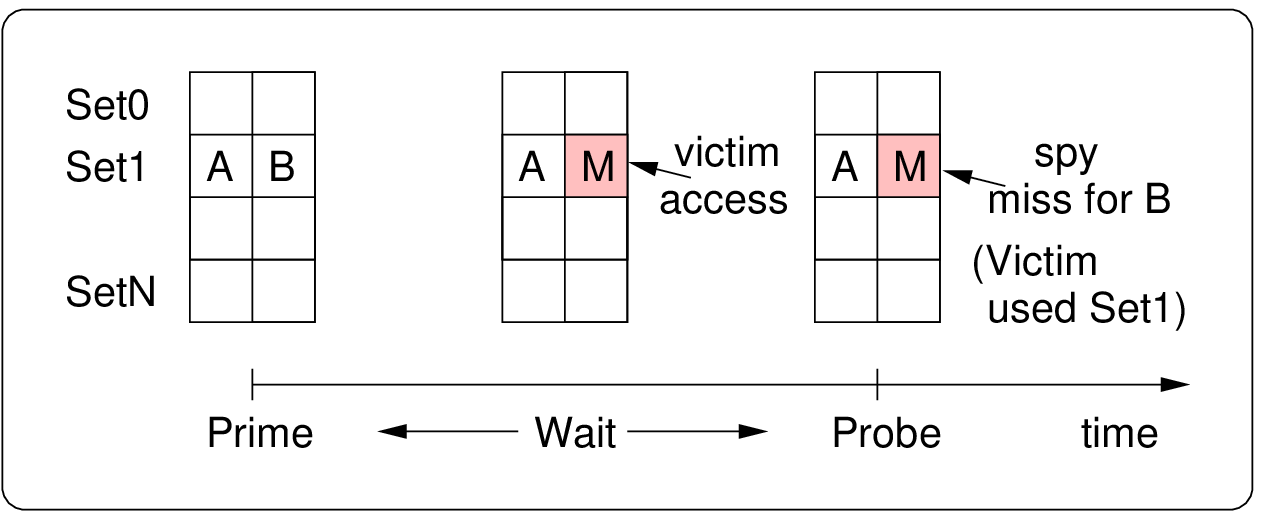}
	\caption{Example of Conflict Based attack (Prime+Probe). Spy uses miss for line B to derive that Victim accessed Set1.}
	\label{fig:conflict} 
\end{figure}
}

\subsection{Conflict-Based Attack and Mitigation}

In conflict-based attacks~\cite{PrimeProbe}, the attacker fills a cache set with its own lines and causes a conflict miss on one of the victim lines.  The attacker uses a timing test to see if any of the installed lines encounter a miss -- if so, the attacker can infer that the victim accessed the particular set.  Fortunately, conflict-based cache attacks can be efficiently mitigated by using cache-space preservation~\cite{RPCache,NoMo, DAWG,SECDCP} or by randomizing the location of the line in the cache~\cite{RPCache,NewCache,CEASER}.  Without loss of generality, in our study, we assume the cache is protected against conflict-based attacks using randomization~\cite{CEASER} and we focus on only flush-based cache attacks.



\subsection{Flush-Based Cache Attack}

Cache attacks do not always have to use load/store instructions to cause cache evictions. They can use an instruction called {\em Cache Line Flush} ({\texttt{clflush}}), which explicitly invalidates the cache line from all levels of the processor cache hierarchy, including the instruction and data~\cite{IntelHandbook}.  The {\texttt{clflush}} instruction is conventionally provided by the system to support non-coherent IO, whereby the IO device can write directly to memory and use {\texttt{clflush}} instructions to ensure that the process can read the up-to-date copy from memory instead of the stale copy from the cache~\cite{IntelHandbook}.

Flush-based attacks target accesses to the memory locations that are shared between the attacker and the victim. In a Flush+Reload attack, the spy invalidates the shared line using the {\texttt{clflush}} instruction, waits, and then checks the timing of a later access to that line -- if the access incurs shorter latency, then the attacker can infer that the victim application has accessed the given line. While Flush-based attacks are restricted to only shared pages, they are more powerful than conflict-based attacks, because the spy can learn about the exact line being used instead of just the particular cache set.  

While users can ensure that data pages containing sensitive data are not explicitly shared with an untrusted application, shared libraries can end up being implicitly shared by the OS. Thus, using library code (like RSA or AES for encryption and decryption) can allow an attacker to monitor the access pattern to different functions and cryptographic tables within those functions, to infer secret information such as cryptographic keys. For example, for the square and multiply routine showed in Figure~\ref{fig:setup}, the adversary can flush the line corresponding to square and multiply functions and learn the access pattern of a victim. We want to develop efficient solutions for tolerating cross-core Flush+Reload attacks.\footnote{Another form of Flush-based attack is the Flush+Flush attack~\cite{FlushFlush}, which exploits the timing difference between flush of a resident line and a non-resident line. We discuss how our solution can be used to mitigate Flush+Flush attacks in Section~\ref{sec:FlushFlushMitigation}.}  

\subsection{Prior Solutions for Flush+Reload Attack}

Current proposals to mitigate Flush+Reload attacks primarily rely on OS-based or software-based techniques. For example, cache attacks on shared data can be avoided by extending the OS to disable sharing of critical pages. Zhou et al.~\cite{CacheBar} proposed a scheme to dynamically replicate shared pages if multiple processes are concurrently accessing such pages, thus giving up on the capacity benefits of page sharing. Gruss et al.~\cite{tmem:usenix2017} proposes to rewrite safety-critical software using transactional memory semantics, which means transactions that have concurrent memory accesses to shared location by other processes will cause transaction abort and avoid leakage of timing information to concurrently running applications.  Prior studies~\cite{PerfCtr1,PerfCtr2} have also suggested using hardware performance counters to observe deviation in the behavior of applications to detect attacks, assuming profile information of the applications (during attack-free scenario) is available. Unfortunately, such profile information may not be available for all applications that use shared pages.  

Architectural solutions for mitigating flush-based attacks fall in two categories. First, redefining the usage of {\texttt{clflush}}. For example, SHARP~\cite{SHARP} proposes to restrict {\texttt{clflush}} in user-mode from flushing read-only pages. Such a solution requires changing the ISA definition of {\texttt{clflush}} and is not backwards compatible. Second, creating in-cache duplicates of the shared line, so a flush triggered by one process cannot dislodge the line brought in by another process. For example, both NewCache~\cite{Ruby:MICRO2017} and DAWG~\cite{DAWG} use in-cache line duplication to mitigate flush-based attack. Unfortunately, such a solution of line duplication is either incompatible with inter-process communication (NewCache) or may require careful placement by the OS to designate the communicating processes within the same security domain (DAWG).

\subsection{Goal of Our Paper}

The goal of this paper is to develop a practical solution to mitigate cross-core Flush+Reload attacks. For a solution to be useful, it is important that it not only provides strong protection against attacks but also has (1) Negligible performance overheads when the system is not under attack, (2) Negligible hardware overhead, (3) No restriction on capacity benefits of page sharing, (4) No requirement of rewriting the software, (5) No changes to the OS or the ISA, and (6) No limitation on inter-process communication. Our paper develops a practical solution to tolerate attacks by leveraging the hardware properties that are inherent in flush-based attacks.  We describe our solution next.

\begin{figure*}[htb] 
  	\centering
  	\vspace{-0.15 in}
	\includegraphics[width=6.75in]{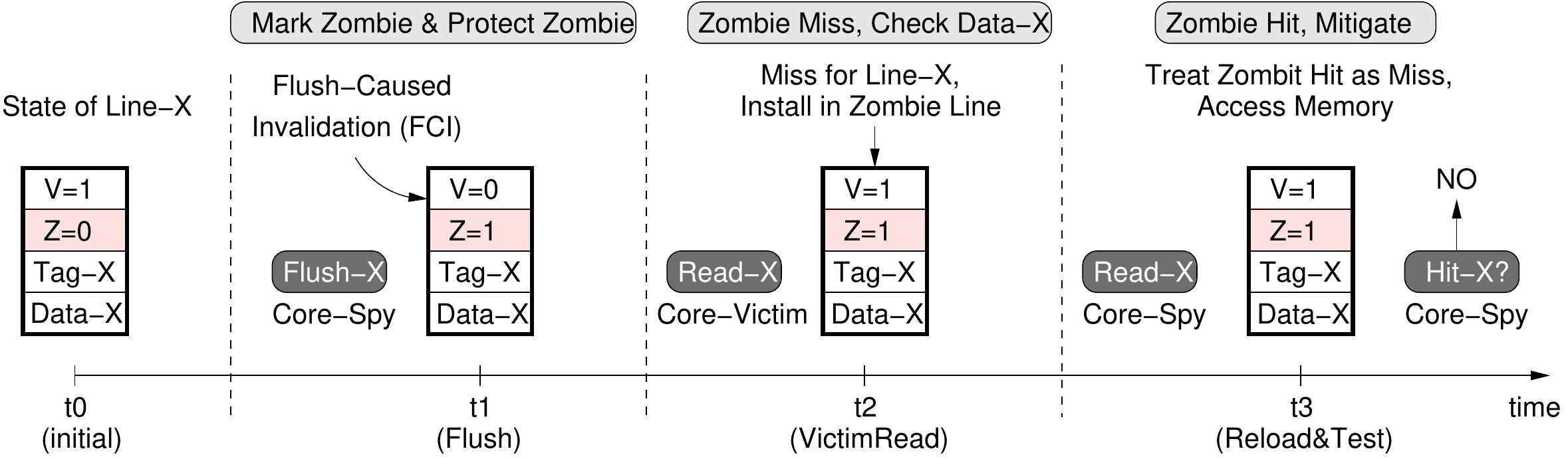}
     \vspace{-0.075 in}
	\caption{Detecting and Mitigating Flush+Reload attack using Zombie. A line that is invalidated due to {\texttt{clflush}} is marked as a zombie (Z=1). A cache miss that has matching tag for a zombie line is termed Zombie-Miss -- incoming line is installed in the Zombie line and Z=1 is maintained if incoming data matches resident data.  A cache hit for a line with Z=1 is termed as Zombie-Hit. Our solution mitigates attack by servicing the Zombie-Hit with a latency of a miss.}
    \vspace{-0.175 in}
	\label{fig:zombie} 
\end{figure*}

\section{Mitigating Attacks via Zombies}
\label{sec:design}
Our objective is to differentiate an ``attack'' from a  benign use-case of {\texttt{clflush}} like non-coherent IO. We observe that for the intended use of {\texttt{clflush}}, the data value resident in memory is expected to change between the flush and subsequent read to the line. For an attack however, a given line is repeatedly flushed, and the data read from memory is unchanged, compared to the data present in the line when it was invalidated due to the flush.\ignore{ And this behavior occurs repeatedly many times.} If we have the line address and data of the lines that were invalidated recently due to a flush,  we can use it to detect Flush+Reload attacks. 

\ignore{
We observe that for the intended use of {\texttt{clflush}} instruction, the data value resident in memory is expected to change between the flush and subsequent read to the line.  For an attack however, a given line is flushed, and the data read from memory is identical to the data that the line contained when it was invalidated due to the flush. And this behavior occurs repeatedly many times. If we have the line address and data of the lines that were invalidated recently due to a flush,  we can use it to detect Flush+Reload attacks. 
}

We leverage the key insight that when a cache line gets invalidated due to a Flush-Caused Invalidation (FCI), the valid bit of the cache line is reset; but the Tag and Data of the invalidated line continue to be present in the cache in the same location. In fact, we can use them for comparison with the Tag and Data of lines installed on subsequent cache misses, to efficiently detect and tolerate Flush+Reload attack. 

\subsection{Determining a Zombie Line}

We call a line invalidated due to Flush-Caused Invalidation a {\em Zombie} line, as it still retains its Tag and Data in the cache. ({\bf Note:} Invalidations due to other reasons such as coherence, system restart, etc. are not deemed Zombies). To track the Zombie status, we add a {\em Z-bit (Zombie bit)} to the tag entry of each line.\ignore{ When a line is installed in the cache, and it is not already present in a Zombie state, the Z-bit is set to 0.} The Z-bit for a line is set to 1, only when it is invalidated due to an FCI. Once set, the Z-bit is reset to 0, only when either the Tag or the Data for the line are updated.

\ignore{
We call the line that is invalidated due to Flush-Caused Invalidation as a {\em Zombie} line ({\bf Note:} invalidations due to other reasons such as coherence, system restart, resizing for power savings, are not deemed as Zombie). To track the Zombie status, we add a {\em Z-bit (Zombie bit)} to the tag entry of each line. When a line is installed in the cache, and it is not already present in a Zombie state, the Z-bit is set to 0. The Z-bit is set to 1 only if the line is invalidated due to an FCI. 
}
\subsection{Basic Operations for Zombie Scheme}

Our Zombie-based design contains four steps: (1) Mark the Zombie (2) Protect the Zombie (3) Actions on Zombie-Miss (4) Actions on Zombie Hits. We explain these steps with an example, as shown in Figure~\ref{fig:zombie}, which is identical to Figure~\ref{fig:intro}, except that Line-X now also contains the Z-bit.


At time t0, the line is resident in the cache with tag as Tag-X and data as Data-X. The valid bit (V) is 1 and the line is not a zombie (Z=0).  At time t1, the line receives a Flush-Caused Invalidation (FCI) from the spy.

\vspace{0.05 in}
\noindent{\bf Step-1.  Mark the Zombie on FCI:} We assume the cache is equipped with signals to identify an invalidation due to {\texttt{clflush}}. On receiving a FCI, the valid bit for the line is reset (V=0) and the line is marked as a zombie line (Z=1). If the line was dirty, the contents are written back to memory (the zombie status is independent of whether the invalidated line was clean or dirty). After the FCI, the line continues to retain Tag-X and Data-X.

\vspace{0.05 in}
\noindent{\bf Step-2.  Protect the Zombie Until Natural Eviction:} Replacement policies preferentially pick invalid lines as replacement victims on a cache miss. So, an invalid zombie line can be quickly dislodged if there is another miss to the same set. To avoid this, we ensure a zombie line resides in the cache until it would have been naturally evicted, in the absence of a FCI. We achieve this by modifying the replacement algorithm to not pick an invalid zombie line as a victim, unless it would have been picked anyway based on its recency/reuse status\ignore{ as per the replacement algorithm}. For example, for LRU replacement policy, the invalid zombie line is not victimized until it becomes the LRU line. All through the residency of the invalid-zombie line in the cache, the line continues to get replacement status updates similar to valid lines (so, for LRU replacement, the invalid zombie line would traverse all the way from MRU to LRU, and only then get evicted). \ignore{ Such protection ensures that the time for which a line remains in the cache is unaffected by FCI. } Protecting invalid zombie lines does not slow down the system as flushes are typically performed only for few (tens of) lines, whereas our L3 cache with randomized indexing~\cite{CEASER} has tens of thousands of locations where new line accesses and installs may potentially map to. 

\ignore{
\begin{tcolorbox}[top=3pt,bottom=3pt]
{\bf Note:} Protecting invalid zombie lines does not slow down the system as flushes are typically performed for only few (tens of) lines, whereas our L3 cache with randomized indexing~\cite{CEASER} has tens of thousands of locations where new lines may be installed.
\end{tcolorbox}

}
\ignore{
\begin{tcolorbox}[top=3pt,bottom=3pt]
{\bf Note:} Given that flush operations are typically performed for only few (tens of) lines and the L3 cache has tens of thousands of lines,  protection of invalid Zombie lines does not impact system
performance.  
\end{tcolorbox}
}
\vspace{0.05 in}
\noindent{\bf Step-3.  Act on Zombie-Miss (tag hit on invalid-zombie):} Conventional cache designs skip the tag match for invalid lines, while probing a set on a cache access. Instead, we perform the tag match even for invalid zombie lines, to identify a cache miss where an invalid zombie line with a matching tag was present in the cache set. We deem such a miss to be a {\em Zombie-Miss}. In the event of such a Zombie-Miss, data is retrieved from memory similar to a normal cache miss. In addition, the incoming cache line is deterministically placed in the cache way where the Zombie line was located, before the valid bit is set. If the data retrieved from memory is different from the data originally in the Zombie line, it indicates a legitimate use of {\texttt{clflush}} (e.g. an asynchronous IO modified the memory contents between the flush and the access), so we reset the zombie status (Z=0) of the line.  However, if the data retrieved from memory is identical to the data originally in the Zombie line, then the Zombie status is retained (Z=1), as the flush was unnecessary and the line could be under attack. This scenario is shown at time t2 in Figure~\ref{fig:zombie}. 

\ignore{
In a conventional cache design, only the valid lines are tested for a tag match. We extend the tag-match logic to also detect cases where there is a invalid-zombie line with a tag identical to what is being accessed.  In such cases, we deem the access as a {\em Zombie-Miss}. Similar to normal misses, data is retrieved from memory on a Zombie-Miss. However, the incoming cache line is deterministically placed in the cache way in which the Zombie-Miss was encountered and the valid bit is set. If the data retrieved from memory was different than what is stored in the cache, it means this was a legitimate use of {\texttt{clflush}} (and that an asynchronous IO modified the memory content between the flush and the access) so we reset the zombie status (Z=0) of the line.  However, if the data retrieved from memory is identical to the data already present in the line, then the Zombie status is maintained (Z=1), as the flush was unnecessary and the line could be a candidate for an attack. This scenario is shown at time t2 in Figure~\ref{fig:zombie}. 
}

\begin{table}[h]
  \begin{center}
    \begin{small}
    \vspace{-0.15 in}
      \caption{Outcome of Cache Access with Zombie  }
      \vspace{0.1 in}
      \begin{tabular}{|c|c|c|c|} \hline

Zombie & Valid & Tag Match? & Meaning and/or Action \\ \hline\hline

X & X & No  & Normal Miss  \\ \hline \hline

0 & 0 & Yes & Normal Miss \\ \hline
0 & 1 & Yes & Normal Hit  \\ \hline
1 & 0 & Yes & Zombie-Miss (Step-3) \\ \hline
1 & 1 & Yes & Zombie Hit (Step-4) \\ \hline

      \end{tabular}
      \label{tab:zombie}
 \end{small}
  \vspace{-0.1 in}
\end{center}
\end{table}

\noindent{\bf Step-4. Act on Zombie-Hit (tag hit on valid-zombie):} On a cache hit, our design checks the Z-bit of the line in parallel with the tag-match. If a cache hit occurs on a valid-zombie line, we deem it as a {\em Zombie Hit}. In a conventional non-secure design, an attacker could use such hits on previously flushed lines and the corresponding short access time, to infer a victim access to the line. Our design can detect such Zombie-Hits and invoke mitigative actions to close the timing leak\ignore{ from access to zombie lines}, as shown at time t3 in Figure~\ref{fig:zombie}.


Table~\ref{tab:zombie} summarizes the cache operations using the zombie-bit in our design. Note that we restrict zombie-bit and associated operations only to the L3 cache, as it is the shared cache in our system, and the one exploited in the cross-core attack.




\subsection{Zombie-Based Mitigation (ZBM)}
Given that Zombie-Misses and Zombie-Hits are intrinsic to the Flush+Reload attack, tracking their episodes can help detect an attack. For example, hardware counters can be maintained that increment on Zombie-Misses and Zombie-Hits, and if the count exceeds a certain threshold, a potential attack can be flagged to the OS to activate mitigating actions.\ignore{ Subsequently, hardware can inform the OS of a possible attack, if these counters exceed a certain threshold.} However, we ideally want a solution that can mitigate the  attack transparently in hardware, without requiring any OS support or threshold-based decisions.  

\ignore{
The episode of Zombie-Hits and Zombie-Misses can be tracked by the hardware to detect and mitigate the attack. For example, dedicated hardware counters can be incremented on Zombie-Hits and Zombie-Miss and the hardware can then inform the OS of a possible attack if these counters exceed a certain threshold. Ideally, we want a solution that can tolerate attacks transparently in hardware, without requiring any OS support or relying on any threshold-based decisions.  
}

Our insight is that the timing difference between a Zombie-Hit and Zombie-Miss is vital for a successful Flush+Reload attack -- eliminating it can completely mitigate the attack. This is because the attack uses timing to guess the outcome of the spy-access following the flush to leak information. A fast access (Zombie-Hit) implies the victim accessed the line between the flush and spy-access, whereas a slow access (Zombie-Miss) implies the victim did not access the line. To eliminate the timing difference between Zombie-Hits and Zombie-Misses and prevent any information leakage, we propose a simple hardware-based scheme, {\em Zombie-Based Mitigation (ZBM)} that delays Zombie-Hits and makes them incur the same latency as a Zombie-Miss.

\begin{figure}[htb] 
  	\centering
	\includegraphics[width=3.30in]{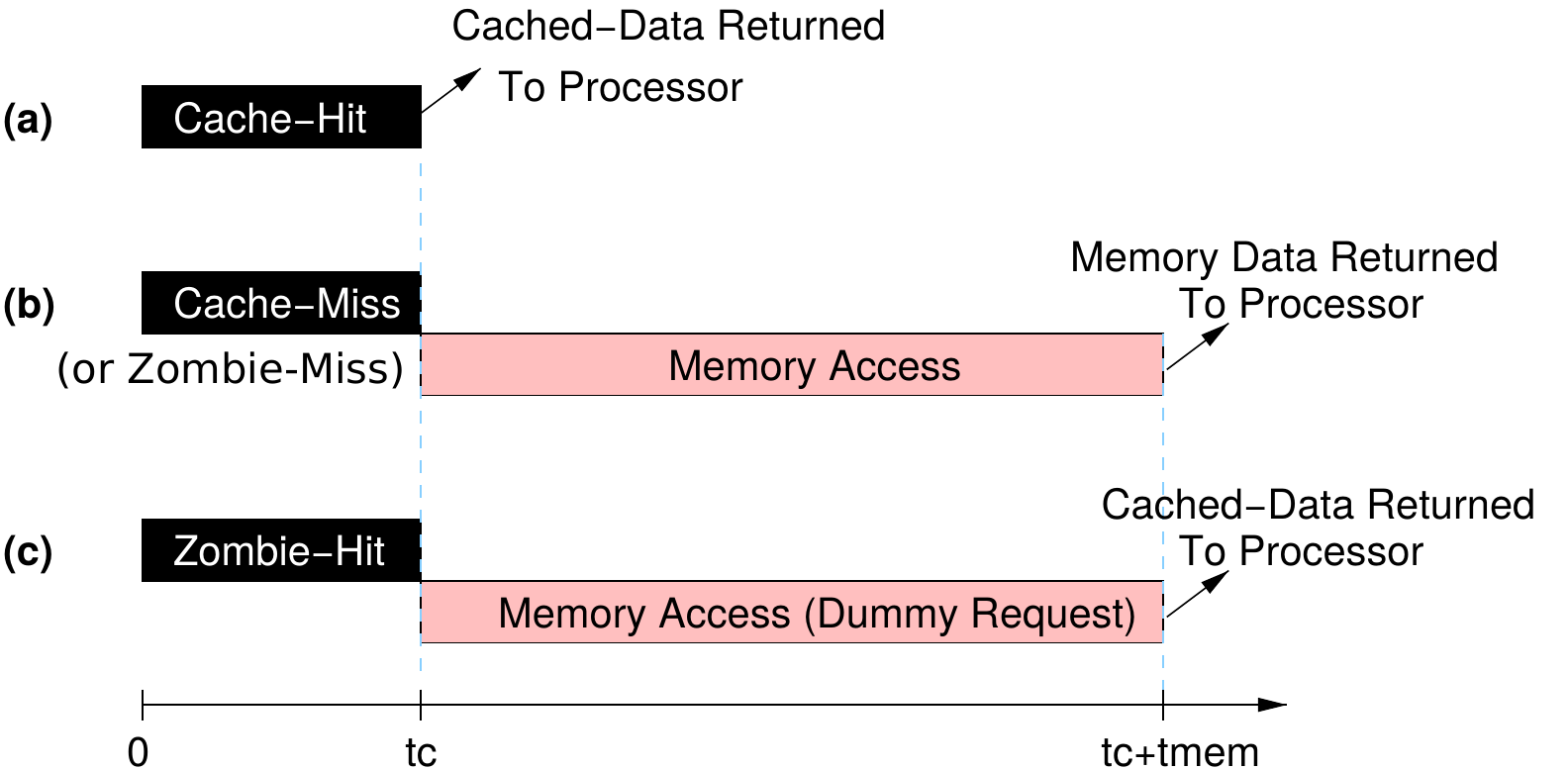}
	\caption{With Zombie-Based Mitigation, timeline for servicing (a) Cache Hit ($t_c$) (b) Cache Miss/Zombie-Miss ($t_c + t_{mem}$) (c) Zombie-Hit ($t_c + t_{mem}$) -- same latency as Zombie-Miss (note that ZBM sends a dummy request to memory to estimate the latency of Zombie-Miss).}
	\vspace{-0.05 in}
	\label{fig:zbm} 
\end{figure}

On Zombie-Hits, ZBM adds delays by triggering an extraneous memory access for the same line, and returning the cached data to the processor only after this dummy request completes (the data obtained from the memory is ignored and not installed in the cache). As this exactly emulates the operations (memory access) on a Zombie-Miss, it incurs a similar long-latency. Thus, during the Flush+Reload attack, the spy-access in the reload phase leaks no information -- the access incurs the high latency of a cache miss and memory access, regardless of whether the line was accessed by the victim (Zombie-Hit) or not (Zombie-Miss). Note that a Zombie-Hit is recorded as a Cache-Miss by the cache performance counters, so no information is leaked even through program statistics. Figure~\ref{fig:zbm} shows the time-line for Zombie-Miss/Hit in our design, illustrating the lack of a timing difference. 

\ignore{With ZBM, when the spy accesses the line in the reload phase of a Flush+Reload attack, the line will incur the high latency of a memory access, regardless of whether the line was accessed by the victim (Zombie-Hit) or not (Zombie-Miss).}

\ignore{
Our insight is that both Zombie-Hits and Zombie-Misses are vital for a successful Flush+Reload attack. If the victim never accesses the line invalidated by the flush, then the spy cannot obtain any information about the victim access pattern. Similarly, for the attack to be successful, there must be a timing difference (hit/miss) for accessing the line after the flush, depending on whether the victim accessed the line, otherwise the spy cannot learn any information about the victim access pattern. We propose a simple hardware-based scheme, {\em Zombie-Based Mitigation (ZBM)} that is based on exploiting these requirements. ZBM mitigates Flush+Reload attack by simply providing a longer latency (equivalent to a memory access) for all Zombie-Hits.  Thus, in the reload phase of a Flush+Reload attack, when the spy accesses the line, the line will always incur high latency regardless of whether the line was absent from the cache (cache miss) or if the victim brought the line in the cache (Zombie-Hit).  On a Zombie-Hit to a line, ZBM triggers an extraneous memory access for  the same line, and returns the cached data to the processor only after this extraneous request from memory is returned (the data obtained from the memory is ignored and not installed in the cache). 
}

The episode of flushing a line, retrieving it with identical content from memory and then accessing it again from the L3 cache and not L1/L2 cache, and doing so repeatedly, does not occur in typical applications. So, the higher latency (and additional energy overhead) due to the extra memory accesses on Zombie-Hits does not impact benign applications.

\ignore{
Figure~\ref{fig:zbm} shows the timeline for a system with ZBM. A cache hit (to a non-zombie line) is served with cache hit latency ($t_c$) and a cache miss is served with an additional memory latency ($t_{mem}$). On a Zombie-Hit the cache waits for a time period of $t_{mem}$ (by triggering and waiting for the extraneous request), so it incurs the same latency as a cache miss. The episode of flushing a line, retrieving it with identical content from memory, accessing it while in the L3 cache, and doing so repeatedly does not occur in typical applications. So, the higher latency (and the additional energy overheads) due to the extra memory accesses on Zombie-Hits does not have any impact for benign applications.
}


\section{Security Analysis of ZBM}
\label{sec:security}
\ignore{
We analyze the effectiveness of  ZBM with three Flush+Reload based side-channel attacks (similar to\cite{tmem:usenix2017}) -- attacking AES T-tables, Square-And-Multiply algorithm of RSA, and Secret-Dependent Function Execution. Additionally, we also evaluate ZBM against a covert-channel attack, where a trojan and a spy communicate using Flush+Reload (similar to \cite{FlushFlush}). We show that ZBM can  successfully mitigate all of these attacks, while incurring negligible overheads. For our evaluations, we use a Pin-based x86 simulator \cite{Pin} running a \hl{8-core system with a 16MB L3 Cache} (system configuration is shown in Table \ref{table:config}) and assume a noise-free setting.
}

We analyze the effectiveness of ZBM at defending against Flush+Reload attack using three representative attacks (adapted from prior works~\cite{SHARP,tmem:usenix2017}):  (1)~Attacking the AES T-tables, (2)~Snooping a victim's control flow  during secret-dependent function execution, and (3) Attacking the Square-And-Multiply algorithm of RSA.  We evaluate the attack on a baseline system without any mitigation and on a system with ZBM. Our system contains 8 cores sharing a 16MB L3 cache (system configuration is shown in Table \ref{table:config}). Despite using a noise-free setting that allows high attack fidelity, we show that ZBM can successfully mitigate all of these attacks.

\subsection{Attacking AES T-Table Implementation}


Commonly used crypto libraries like OpenSSL and GnuPG implement AES with T-tables, which have been shown to be vulnerable to cache-attacks that leak the secret key \cite{Bernstein,tmem:usenix2017,Bonneau}. In such an implementation, in each of the 10 AES rounds, a total of 16 accesses are made to T-tables (lookup tables) spread over 16 cachelines. The table indices accessed in each round depend on the input plaintext $p$ and a secret key $k$, with the first-round indices  being  ($x_i = p_i \bigoplus k_i $), where $i$ is the byte number in the plaintext or the key. 

We perform a chosen-plaintext attack on the first round. In such an attack, the spy supplies a series of plaintext blocks to be encrypted by the victim,  keeping byte $p_0$ constant and randomly varying the other $p_i$ for  $i\neq 0$. This ensures that a fixed index $x_0$ always receives a hit in the first round, and has a higher number of hits on average compared to other entries in the T-table. By identifying which T-table cacheline received maximum hits using a Flush+Reload attack over a large number of encryptions, the spy can decipher 4-bits out of 8-bits of the index $x_0$ (that determine the cacheline number) and thus extract 4 out of 8 bits of $k_0$ (as $p_0$ is known). By sequentially changing which  plaintext byte is constant, the spy can recover 64 out of 128 bits of the secret key. 

\begin{figure}[htb] 
    \vspace{-0.075in}
  	\centering
	\includegraphics[width=3.50in]{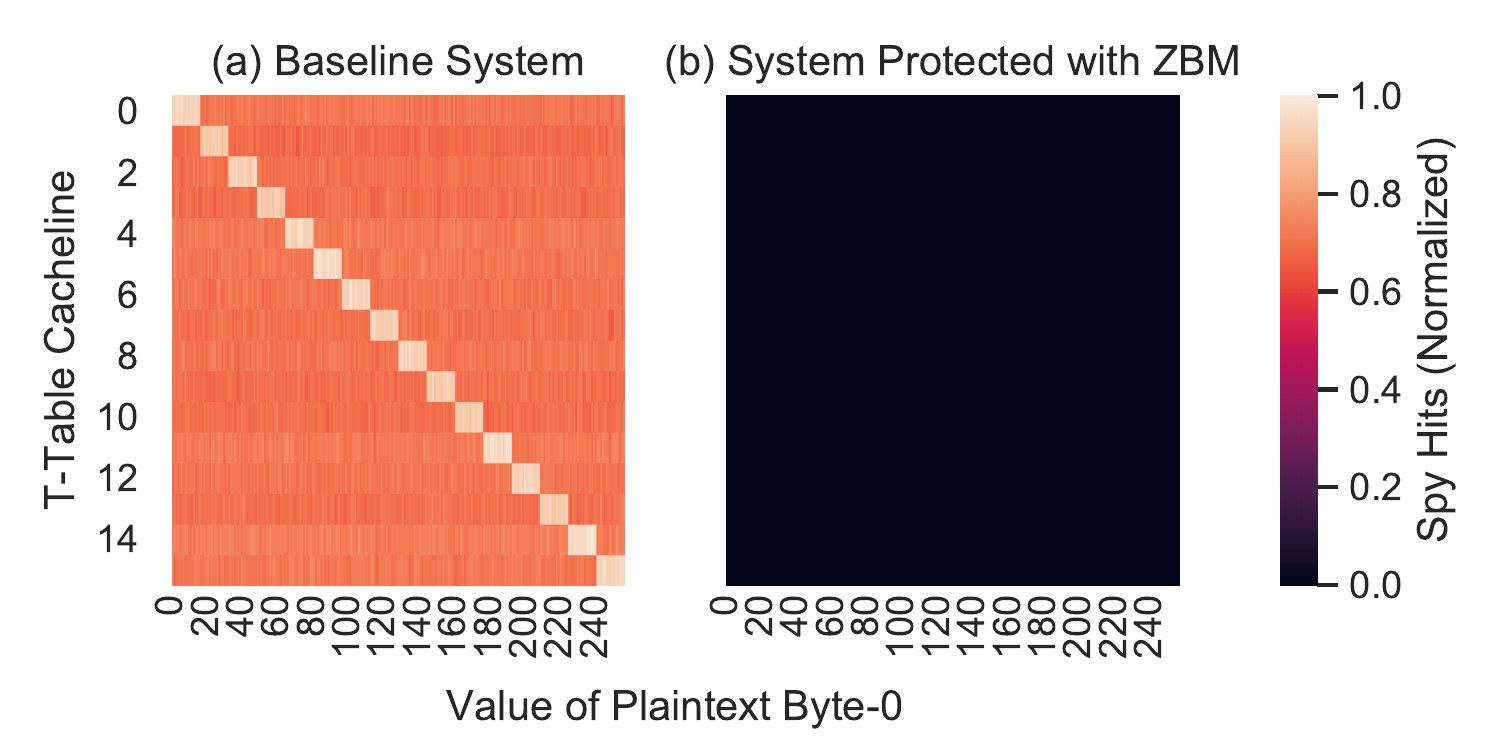}
    \vspace{-0.2in}
	\caption{Flush+Reload Attack on AES T-tables for the (1) baseline system without ZBM and (b) system with ZBM protection. Heat map of hits to T-table cachelines seen by spy, for different values of plaintext byte $p_0$, when the first byte of the key $k_0=0$.}
	\label{fig:aes} 
\end{figure}

\begin{figure*}[htb] 
  	\centering
	\includegraphics[width=7.0in]{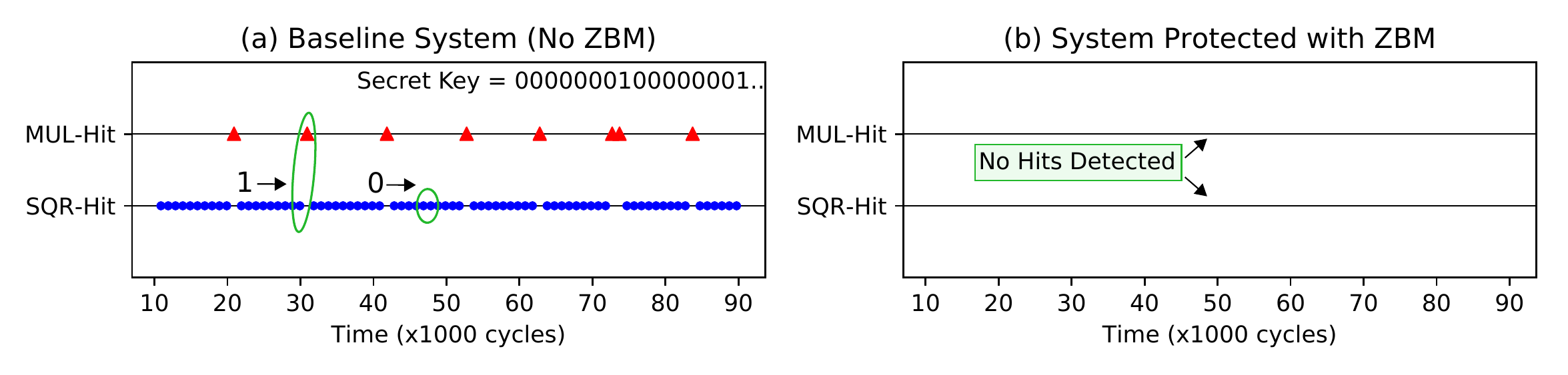}
	\vspace{-0.25in}
    \caption{RSA Attack: Time-line of spy-hits for cachelines in \textit{sqr} and \textit{mul} functions in (a) Baseline (b) ZBM. Here, a  SQR-Hit followed by MUL-Hit leaks bit-value "1" in the key, whereas two SQR-Hits one after another denote a "0".}
	\label{fig:rsatime} 
	\vspace{-0.075in}
\end{figure*}

Figure~\ref{fig:aes} shows the number of hits observed by the spy across the cachelines of the T-tables for different values of first plaintext byte $p_0$, with 10K blocks encrypted per plaintext. The results for both the (a) baseline system without ZBM and (b) with ZBM protection are normalized to the maximum hits for any cacheline across both configurations. 
In the baseline, as $p_0 $ (and $x_0 = p_0 \bigoplus 0$)  increases from 0 to 255, the cacheline corresponding to $x_0$ (the table index with the maximum hits) discernibly changes from 0 to 15 as seen in Figure~\ref{fig:aes},  with each cacheline storing 16 entries. Compared to the average of 1368 hits/cacheline, the maximum observed  is 1892 hits/cacheline. With ZBM,  the spy does not see any pattern because ZBM always provides data with latency of memory access on both Zombie-Hits and Zombie-Misses.\footnote{In a few rare cases, we see a single hit for the spy across all the 10,000 block encryptions. This happens when a flush from the spy occurs before the victim had a chance to access the table even a single time. So on the flush, the Z-bit is not set as the cache does not even contain the line. Subsequent victim access followed by the spy access causes the cache hit for the spy. In the AES attack, all sixteen lines of the tables are accessed thousands of times, with varying probabilities. Knowing one line was accessed once by the victim at the start, is not meaningful information for the attacker.}

\subsection{Function Watcher - Leak Execution Path}

    Just as a spy can snoop the data-access pattern of the victim, it can also snoop a victim's code-access patterns to leak information. Recently, such attacks were demonstrated on PDF processing applications~\cite{Hornby2016} where a spy snoops the functions executed by the program to identify the rendered PDF. On these lines, we model a {\em Function Watcher} attack similar to prior work~\cite{tmem:usenix2017}, where a victim executes a call to one out of four functions based on a secret value (each function has 5000+ instructions with varying control-flow). The spy  monitors the addresses corresponding to the entry points of these functions by repeatedly executing a cross-core Flush+Reload attack. On each reload following a flush, the spy observes which address received a cache-hit (lowest latency), to infer the function being executed by the victim.

\begin{figure}[htb] 
  	\centering
	\includegraphics[width=3.55in]{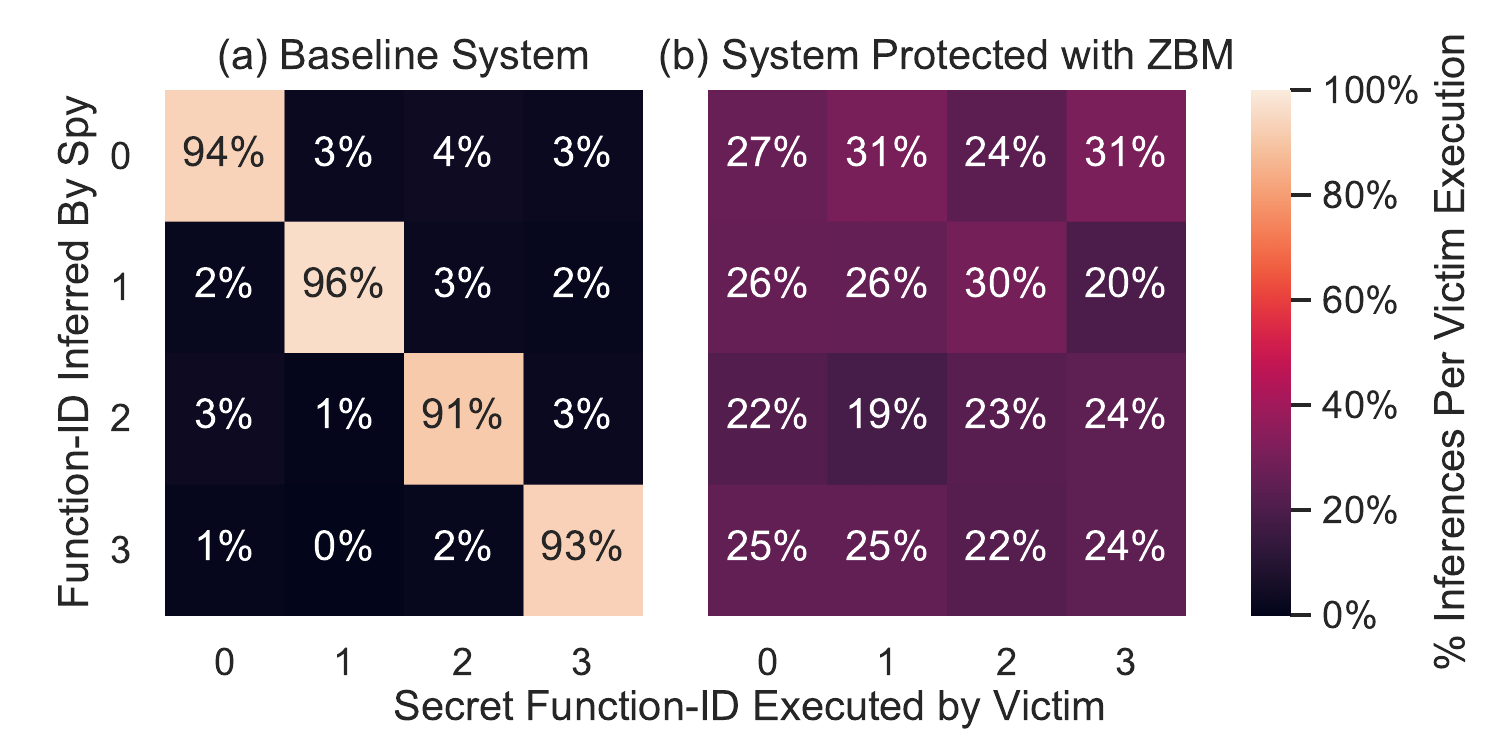}
   \vspace{-0.2in}

	\caption{Function Watcher - A spy inferring which one out of four functions was accessed by a victim. Heat-map shows percentage of attempts where a Function-ID $x$ executed by the victim was inferred as $y$ by the spy. In (a)~Baseline the secret can be correctly inferred by the spy with >90\% accuracy, while in (b)~ZBM the attack is only as good as a random-guess.}
	\label{fig:funcwatch} 
\end{figure}

The heat-map in Figure~\ref{fig:funcwatch} shows the percentage of attempts where the spy correctly (diagonal-entries) or incorrectly (non-diagonal entries) infers the function executed by the victim, over 10K function calls. In the baseline (Figure~\ref{fig:funcwatch}(a)), the spy correctly infers the function ID of the victim (the secret) in most of the attempts (91\% - 96\%).

In ZBM (Figure~\ref{fig:funcwatch}(b)), the attacker successfully infers the correct function ID in only 23\% - 27\% of the attempts, which is as likely as other incorrect function IDs (19\% - 31\%). Thus, the attack with ZBM is no better than a random guess. This is because ZBM ensures that on a spy-reload, both Zombie-Hits (function entry-point accessed by victim) and Zombie-Misses (unaccessed addresses) have similar high latency, thus leaking no information to the spy.

\begin{figure*}[htb] 
  	\centering
   	\includegraphics[width=7in]{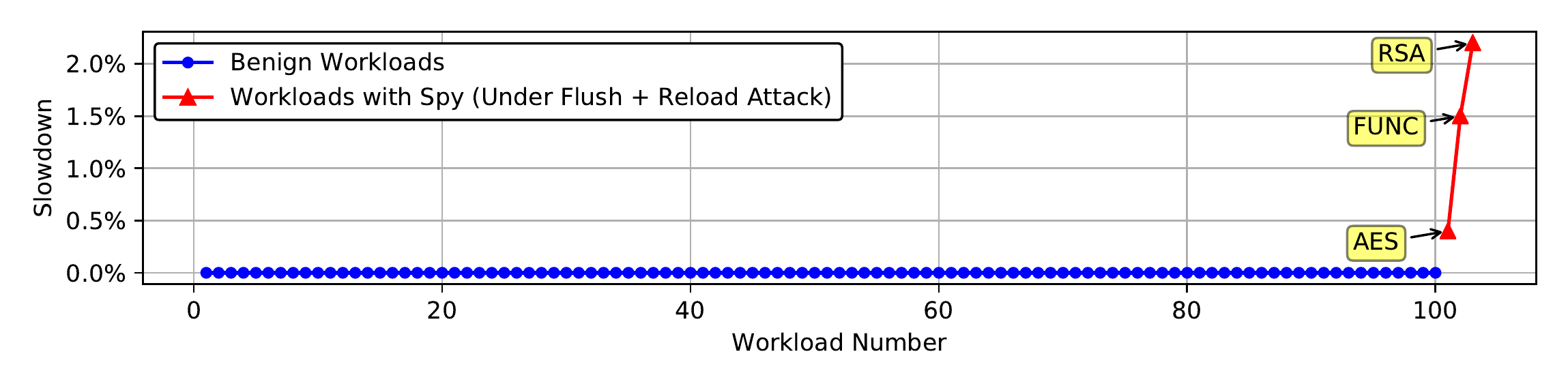}
   \vspace{-0.2 in}
	\caption{Normalized system performance with ZBM for 100 random mixes of benign applications + 3 cases of attack (victim and spy). ZBM incurs no slowdown for benign workloads and also has negligible overheads under attacks.}
	\label{fig:perf} 
\end{figure*}

\subsection{RSA Square-And-Multiply Algorithm}

To evaluate the effectiveness of ZBM, we analyze an attack on the Square-And-Multiply algorithm in RSA described in Section \ref{sec:rsa_bg}, similar to prior work~\cite{SHARP}. In this attack, the spy monitors the victim's accesses to the entry-points of square function (SQR) and multiply function (MUL), to extract the secret RSA key. Figure~\ref{fig:rsatime} shows the time at which spy-hits for SQR and MUL are encountered, in a time window of 10K - 90K cycles for both the baseline (left) and ZBM (right). 

\newpage

In our analysis, we use a RSA key of 3072 bits where every 8th bit is a 1 (e.g key is 0000000100000001...). For the attack in the baseline, a hit to SQR followed by a hit to MUL indicates a "1" in the key, whereas a hit to SQR followed by another hit to SQR denotes a "0". So we expect the spy would have eight hits to SQR followed by one hit to MUL. In Figure~\ref{fig:rsatime}(a), the spy can observe this exact pattern after filtering noise (e.g.a spurious double hit for MUL at time 73K cycles is ignored, knowing the behavior of RSA). With ZBM, the spy always gets high latency on a reload for MUL and SQR -- so the spy does not get any hits, as shown in Figure~\ref{fig:rsatime}(b). Thus, ZBM successfully mitigates the attack.

\ignore{

\begin{figure}[htb] 
  	\centering
	\includegraphics[width=3.350in]{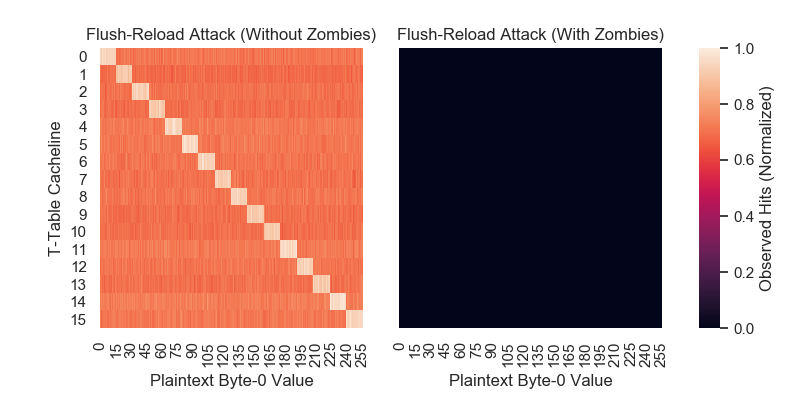}
	\caption{RSA timeline of hits for the spy for lines corresponding to sqr and mul for (a) baseline (b) ZBM}
	\label{fig:rsa} 
\end{figure}

}

\section{Performance Evaluation}
\label{sec:perf}

\subsection{Configuration}

For our performance evaluations, we use a Pin-based x86 simulator. Table~\ref{table:config} shows the system configuration used in our study. Our system contains 8 cores, each of which have a private instruction and data cache and a unified 256KB L2 cache. The L3 cache is 16MB and is shared between all the cores. All caches use a line size of 64 bytes.

\begin{table}[h]
  \begin{center}
    \begin{small}
  \vspace{-0.15 in}
      \caption{Baseline Configuration  }
      \begin{tabular}{|l|l|} \hline

\multicolumn{2}{|c|}{\bf Processor} \\ \hline

Core parameters      &   8-cores, 3.2GHz   \\ 
L1 and L2 cache      &  32KB, 256KB 8-way (private) \\ \hline

\multicolumn{2}{|c|}{\bf Last Level Cache} \\ \hline

L3 (shared)          &   16MB, 16-way, SRRIP, 24 cycles \\ \hline

\multicolumn{2}{|c|}{\bf DRAM Memory-System} \\ \hline

Bus frequency        &    1200 MHz (DDR 2.4 GHz)  \\
Channels             &      2  (8-Banks each)        \\
tCAS-tRCD-tRP   & 14-14-14 nanoseconds\\ \hline

      \end{tabular}
      \label{table:config}
 \end{small}
\end{center}
  \vspace{-0.15 in}
\end{table}

We measure aggregate performance of our 8-core system using weighted speedup metric. We evaluate performance under two scenarios: (1) when the system is not under attack, executing only benign applications (common case), and (2)~when system is under attack (uncommon case). 

For benign applications, we use a representative slice of 250 million instruction of each of the 29 benchmarks in the SPEC2006 suite. The benign applications do not have a "Flush+Reload" pattern. We create a benign workload by randomly pairing 8 benign applications, and have 100 such mixed workloads. For attacks, we use 3 pairs of victim-spy (AES, Function Watcher and RSA) and combine each victim-spy with six benign applications to create an 8-core workload. 

\subsection{Impact on Aggregate System Performance}

ZBM inflicts higher latency for a Zombie-Hit. For such scenarios to occur the line must be flushed, reloaded with identical content soon (before the line gets evicted from the cache) and then accessed again while the line is in the L3 cache. Such access patterns are extremely unlikely to occur in normal (benign) applications, therefore the impact of ZBM on system performance for benign workloads is expected to be negligible.  Figure~\ref{fig:perf} shows the slowdown caused by ZBM for 103 workloads (100 benign + 3 under attack). For calculating slowdown we compute the ratio of weighted speedup of the baseline to the weighted speedup of ZBM. For all 100 workloads that do not have an attack, ZBM has no slowdown, whereas there is a slowdown (up-to 2.2\%) for the three attack workloads containing the victim-spy pairs. 


\subsection{Slowdown for Applications Under Attack}

\begin{figure}[htb] 
  	\centering
   	\includegraphics[width=3.5in]{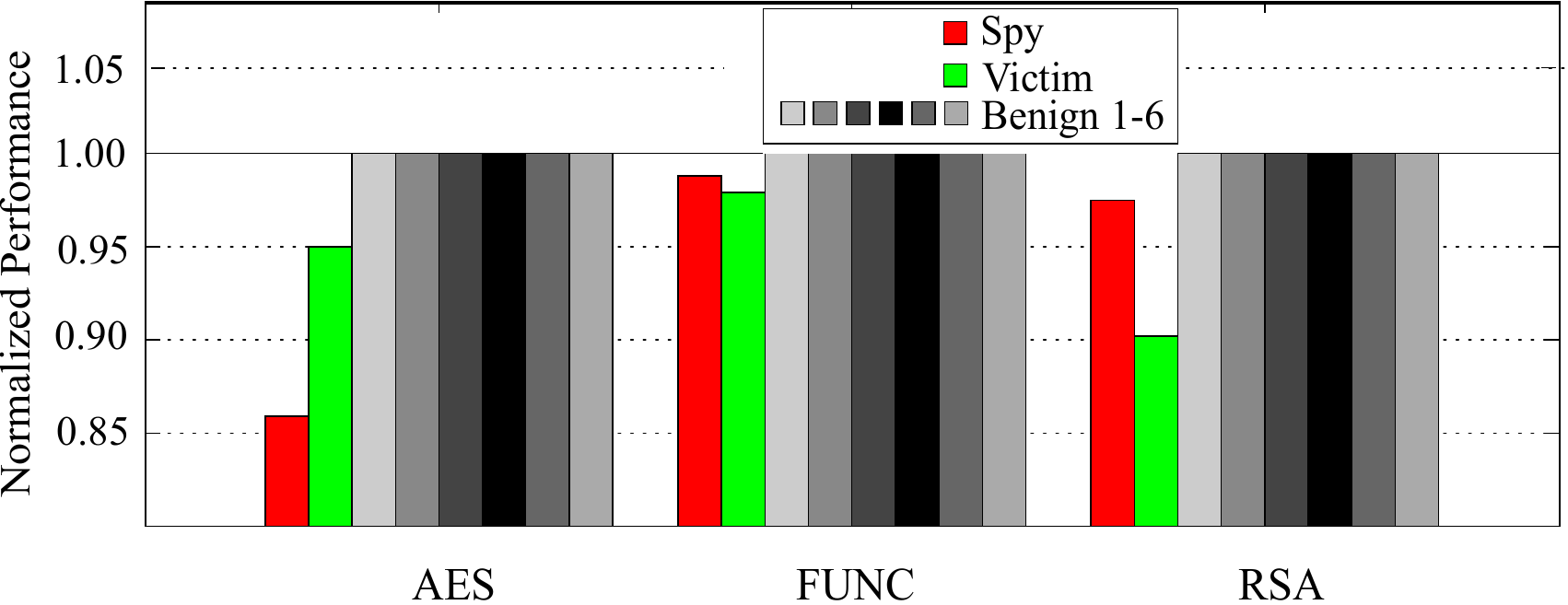}
	\caption{Application-wise performance for the three workloads with victim-spy. ZBM causes slowdown for only the victim and spy, but not the benign co-runners.}
	\label{fig:indperf} 
\end{figure}

We analyze the performance of the individual applications for the three attack  workloads that experience a slowdown. Figure~\ref{fig:indperf} shows the slowdown for each of the eight  applications in the three workloads.  The workload contains a spy, a victim, and six benign bystanders. For AES, ZBM causes 16.4\% slowdown for the victim and 5.3\% for the spy. For Function watcher, the slowdown is 1.2\% for the victim and 2.1\% for the spy. For RSA, the slowdown is 2.6\% for the victim and 10.9\% for the spy. Note that these slowdowns occur only under attack (when marginal slowdown is acceptable to protect against the attack).  The bystander applications (marked as Benign 1-6 in Figure~\ref{fig:indperf}), which are neither a victim nor a spy do not see a performance impact.

\subsection{Storage and Logic Overheads }

ZBM provides strong protection against attack while causing no performance overhead when the system is not under attack. To implement ZBM, the only storage overhead incurred is the per-line Z-bit. Therefore, to implement ZBM on our 16MB LLC, we would need a total storage overhead of 32 kilobytes, which is less than 0.2\% of the LLC capacity.


ZBM requires that the incoming data from memory be compared to the data resident in the cache, if the Z-bit associated with the victim line is set. To implement a 512-bit comparator, we need 512 XNOR gates and 511 AND gates, for a total of approximately 1K gates. This logic overhead is negligibly small (for reference, computing ECC-1 for a 64-byte line incurs an overhead of approximately 3500 gates).



\section{ZBM-X for Flushing Applications}

\begin{figure*}[htb] 
  	\centering
   \vspace{-0.35 in}
   	\includegraphics[width=6.3in]{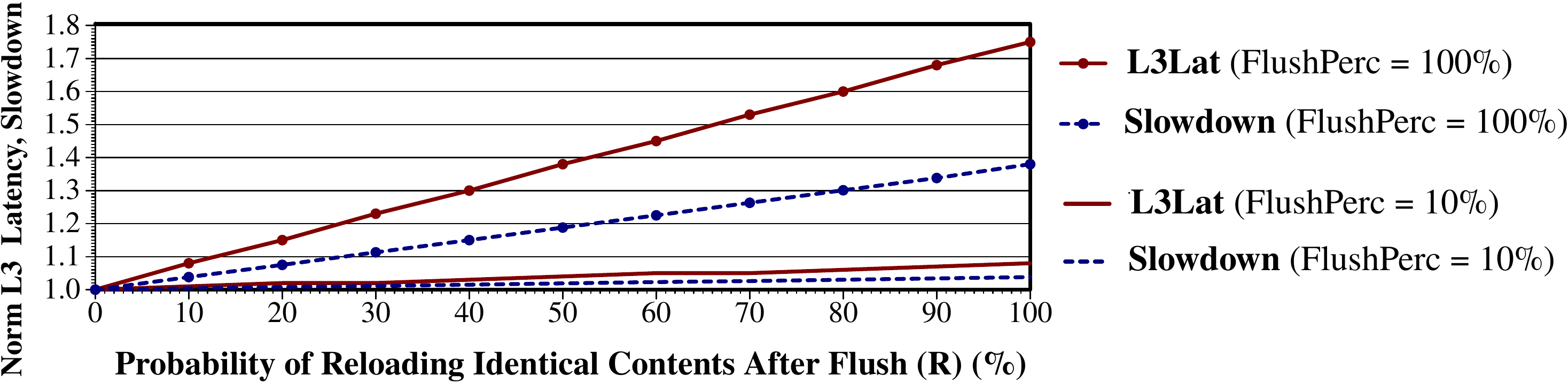}
	\caption{Normalized Average L3 Access Latency (L3Lat) and Slowdown with ZBM, as a function of Flush-Percentage (F) and Probability of Reloading Identical Content (R). ZBM slowdown is negligible, if either F or R are small. Even for pathological scenarios with Flush and Reload of Identical Data all the time (F=100\%,R=100\%), the slowdown is <40\%. }    
	\label{fig:amodel} 
\end{figure*}

Our performance evaluations focus on workloads that are attacker-victim pairs (demonstrating the security of ZBM). The typical benign workloads we evaluate\footnote{Typical use of {\texttt{clflush}} is for non-coherent IO~\cite{IntelHandbook} -- such applications are not amenable to typical tracing infrastructure, so tracing them is beyond our scope. Instead, we develop an analytical model to estimate slowdown of ZBM for such flushing applications.} do not contain patterns of Flush+Reload of identical data, and hence have no slowdown with ZBM. However, there could be other niche applications which use flushes frequently. For example, lines evicted with \texttt{clflush} in expectation of a non-coherent IO, where no IO occurs, could be read subsequently as Zombie Hits with ZBM. Other unintended usage of \texttt{clflush} can occur in persistent memory applications, which use it to evict and write dirty cache lines to memory, even though more suitable instructions like \texttt{clwb} (cache line writeback) exist.

Even with frequent flushes, ZBM's slowdown is expected to be minimal as several conditions need to be true at the same time to suffer slowdown -- (a)~application has to flush lines, (b)~then reload identical data, and (c)~the data has to be larger than private L2-cache (256 KB), so that subsequent loads to it are all serviced from L3-cache. This is uncommon, as subsequent loads are likely to be serviced, in most scenarios, from L1 or L2 cache without overheads (Z-bit is restricted to the L3 cache). Nonetheless, to understand the slowdown of ZBM for flushing applications, we develop an analytical model. Then, we design a simple extension of ZBM to avoid slowdown even in the uncommon scenario of frequent flushes.

\ignore{
Our performance evaluations focus on workloads that are attacker-victim pairs (to demonstrate the security of ZBM) and do not contain workloads with Flush+Reload of identical data.\footnote{The typical use of {\texttt{clflush}} is for non-coherent IO devices~\cite{IntelHandbook} (modern systems typically use a coherent DMA, so DMA transfers are not required to use {\texttt{clflush}}). It is beyond the scope of our paper to trace the data of non-coherent IO devices, as such devices are not typically amenable to the tracing infrastructure used in the analysis of regular applications. Therefore, we develop an analytical model to estimate the slowdown of ZBM on such flushing applications.} However, there could be scenarios where an applications that uses flushes can get impacted by ZBM. For ZBM to cause slowdown, several conditions need to be true at the same time-- in particular, (a) application flushes a line, and (b) then reloads the line with the same data, and moreover (c) the data does not fit in the private L2-cache (256 KB).  Even if a benign program flushes a line, then reloads identical data, setting the Zombie-bit in L3-Cache, subsequent loads would be serviced from L1/L2 without any overheads. Nonetheless, to understand the performance impact of ZBM on flushing applications, we develop an analytical model. Then, we develop a simple extension of ZBM that can avoid the performance impact even for such uncommon applications.
}

\subsection{Analytical Model - L3Lat and Slowdown}

Slowdown with ZBM is due to Zombie-Hits being serviced with the latency of a L3-cache miss instead of a L3-cache hit, to prevent leaking information. So, we first develop a model to estimate the \textit{average latency for a L3 access} (\textit{L3Lat}) with ZBM and then model its impact on execution time.

\ignore{
As ZBM services hits to the Zombie lines with the latency of a cache miss to prevent leaking information, the slowdown of ZBM comes from such extra misses. Therefore, we first develop a model to estimate the impact of ZBM on the {\em Average Memory Access Time (AMAT)} of the L3-cache. 
}

\vspace{0.02in}
\subsubsection{Modelling L3 Access Latency (L3Lat)}
Let $\alpha$ be the miss-rate of the baseline L3-cache.  Let $t_c$ be the number of cycles for a L3-cache hit and $t_m$ be the number of cycles to service a memory access. Then \textit{L3Lat} for the baseline system ($L3Lat_{base})$ is given by Equation 1.

\vspace{-0.01in}
\begin{equation}
L3Lat_{base} = t_c +  \alpha \cdot t_m 
\end{equation}

\vspace{-0.075in}
Let $F$ be the percentage of memory accesses that occurred due to a flush of the cache line (line was present in the cache but got invalidated due to clflush). Let $R$ be the probability that the reloaded content is identical to the content of the line at the time of the flush.  Thus, $F\cdot R$ represents the probability of flush and reload of identical content.  If all memory accesses were flush-reload of identical contents (i.e. $F\cdot R = 1$), then we expect the L3-miss rate in ZBM to increase to 100\%. 

\vspace{0.05in}
In general, for any given value of F and R, the miss-rate of the ZBM ($\alpha_{ZBM}$)  is given by Equation 2.

\vspace{-0.025in}
\begin{equation}
\alpha_{ZBM} = \alpha + (1-\alpha)\cdot F \cdot R
\end{equation}

\vspace{-0.025in}
Using Equation 1 and 2, the \textit{L3Lat} for the ZBM system ($L3Lat_{ZBM})$ is given by Equation 3.

\begin{equation}
L3Lat_{ZBM} = t_c + (\alpha + (1-\alpha)\cdot F \cdot R) \cdot  t_m 
\end{equation}

Dividing Equation 3 by Equation 2, we get the normalized value ($L3Lat_{norm}$), as shown in Equation 4 and in Figure~\ref{fig:amodel}.

\begin{equation}
L3Lat_{norm} = \frac{t_c +  (\alpha + (1-\alpha)\cdot F \cdot R) \cdot t_m }{t_c +  \alpha \cdot t_m} 
\end{equation}

\vspace{0.02in}
\subsubsection{Modelling Overall System Slowdown}
Execution time of the baseline system ($T_{base}$) can be split into two components, the time required with a perfect L2-cache ($T_{perfL2}$) and the time spent accessing L3 and memory ($T_{L3mem}$), as shown in Equation 5.

\begin{equation}
T_{base} = T_{perfL2} + T_{L3mem}
\end{equation}

\newpage
For ZBM, $T_{L3mem}$ increases in proportion to \textit{L3Lat}. Therefore, the execution time of the application with ZBM ($T_{ZBM}$) can be written as Equation 6.

\begin{equation}
T_{ZBM} = T_{perfL2} + L3Lat_{norm} \cdot T_{L3mem}
\end{equation}
\vspace{-0.2in}

The slowdown of ZBM can be obtained by dividing Equation 6 by Equation 5, as shown in Equation 7.

\begin{equation}
Slowdown = 1 + (L3Lat_{norm}-1) \cdot \frac{T_{L3mem}}{T_{base}}
\end{equation}
\vspace{-0.1in}

\vspace{-0.1 in}

\subsection{Impact on L3Lat and Slowdown}

\ignore{We use Equation 4 and Equation 7 to analyze the impact of ZBM on a generic flushing application.  }For our system, we use $t_c$ = 24 cycles and $t_{m}$  = 145 cycles. Without loss of generality, we assume an application that spends half of the execution time in L3/memory accesses ($T_{base} = 2 \cdot T_{L3mem}$) and has a baseline L3 hit-rate ($\alpha$) of 50\%. Using these values with Equations 4 and 7, we plot L3Lat and Slowdown in Figure~\ref{fig:amodel} for different values of F and R. Our model shows that an application reloading identical data after flush for <10\% of its accesses will have negligible slowdown with ZBM. Moreover, even pathological scenarios where all memory accesses are reloads of unchanged data after flush, have a  slowdown that is less than 40\%.

\subsection{ZBMx (an extension) to Avoid Slowdown}

It is expected that only an uncommon class of workloads incur slowdown with ZBM -- with frequent flushes to majority of their working set, reload of identical data from memory to cache, and a high hit-rate in L3 for the reloaded contents. To avoid slowdown even in such uncommon scenarios, we extend ZBM to allow cache-hits on a Zombie-line, once the line had a reload from the flush-causing-core, that nullifies the effect of the flush. We call this design extension \textit{ZBMx}.

\ignore{\footnote{Even for persistent memory implementations, {\texttt{clwb}} (cache line writeback)  is preferred over the {\texttt{clflush}} instruction~\cite{Moneta}. This is because {\texttt{clwb}} has better ordering properties (ordered at SFENCE) than {\texttt{clflush}} (ordered at MFENCE) and {\texttt{clwb}} performs the writeback of the cache line without invalidating the cache contents. Nonetheless, if a persistent memory application uses {\texttt{clflush}} and reloads identical contents,  and has a high hit-rate for such reloaded data in the L3-cache (while having low hit-rate for such data in the L2 cache), then we can avoid the slowdown of ZBM using ZBMx.}
If slowdown for such workloads is still a concern, it can easily be addressed by a simple extension of ZBM that avoids the slowdown when flush and refill (cache line install) happens from the same core. We call such an extension of ZBM as {\em ZBMx}.
}

ZBMx is implemented by tracking the {\em Flush-Causing Core-ID (FCID)}, which is set along with the Z-bit, when the line is invalidated due to a {\texttt{clflush}}. When a subsequent access occurs from the same core (coreID matches the line's FCID), then the Z-bit of the line is reset. In this design, subsequent accesses after a Flush and Reload (by the same core) get the line with cache-hit latency, without any slowdown.
\ignore{
ZBMx avoids slowdown of ZBM even for such pathological applications, as the flush and the reload of the identical content would happen from the same core.
}

ZBMx incurs a storage overhead of 4 bits per line  (Z-bit + 3-bit of FCID for an 8-core system), which is still less than 1\% of the area of the L3-cache. Thus, system designers can implement ZBMx with minimal storage overhead, to avoid slowdown even for applications with frequent flushes like non-coherent IO, DMA or persistent memory applications where ZBM might be a potential concern.


\newpage

\section{Security Discussion}
\label{sec:attacks}

Here, we discuss how ZBM tolerates other attack variants\ignore{, given that Flush+Reload attack is successfully mitigated}.

\subsection{Mitigating Flush+Flush Attack}
\label{sec:FlushFlushMitigation}
Flush+Flush~\cite{FlushFlush} attack exploits the variation in latency of the {\texttt{clflush}} instruction depending on whether or not the line is present in the cache, to infer the access of a victim. Such attacks can be prevented by always serving a flush with a constant worst-case latency. To limit slowdown to only a potential attack scenario, we can activate such constant-time service of flush only if the flush is to a line with the Z-bit set -- as repeatedly executing a successful attack requires performing a flush on a recently flushed line (with Z-bit set).

\subsection{Tolerating Alternate Eviction Mechanisms}
\textbf{Cache Conflicts:} Cache-set conflict based attacks, like Prime+Probe~\cite{PrimeProbe}, can be prevented by using way partitioning~\cite{NoMo,DAWG} or by randomizing the cache indexing~\cite{CEASER,NewCache,RPCache}. We assume our system has randomized indexing~\cite{CEASER} for its shared cache and directory~\cite{DirectoryAttacks}.  

\textbf{Natural Eviction or Cache-Thrashing:} An adversary could wait for the victim's sensitive lines to get naturally evicted (e.g. by reaching LRU position in a set), through the victim's own accesses. Alternately, the adversary could attempt to evict all lines including Zombies by cache-thrashing (e.g. by accessing a large array). For conclusively evicting a zombie line in either case, with a randomized-indexing cache, the adversary would need to wait for tens of thousands of lines to be installed (to replace the entire contents of the cache). Such an attack is 10,000x slower than Flush+Reload, that needs just one access for eviction, and much less practical.

\textbf{Non-Temporal Stores (NT-Stores)} can also evict a cached line while writing to memory. However, this can be avoided with an implementation  (like Pentium-M~\cite{IntelHandbook}) that updates a line in-place in the cache without eviction, if the location is cached. This leaves common-case usage of NT-Stores unchanged, where stores directly write to memory when the location is uncached,\footnote{NT-Store should also update the data for a invalid zombie-line if present in the cache (a presence check is anyway needed). Thus, an adversary cannot reset the Z-bit with NT-store and subsequent load.} without costly Read-For-Ownership.

\subsection{Tolerating Alternate Reload Strategies}
Alternate attack variants such as Flush+Prefetch~\cite{EvictPrefetch} and Invalidate+Transfer~\cite{CoherenceInvalidate} perform the reload through a prefetch or a coherence operation, instead of a direct demand access. ZBM tolerates these attacks, as it always services zombie-hits on reload (regardless of its cause) with cache-miss latency.

\subsection{Implications for Denial of Service (DoS)}
An adversary can attempt a DoS attack by flushing shared lines and creating a large number of zombies in the cache. Subsequent victim accesses (zombie-hits) will have slowdown, until zombie lines are naturally evicted. Fortunately, this only impacts the L3 cache, as the zombie status is not propagated to L1/L2 cache -- the application retains use of L1/L2 caches. Worse DoS-attacks not involving ZBM are possible even in the baseline, where an adversary can repeatedly flush lines from all levels of the cache hierarchy, disable the use of all caches and cause much higher slowdown.\ignore{ for the victim resulting in much higher slowdown. }

\subsection{Tolerating Main-Memory Attacks}
While our primary goal is mitigating Flush+Reload cache attacks using Zombie-lines, we observe that Zombies can also help in detecting other attacks that require repeated use of \texttt{clflush}. In this section, we discuss the application of Zombie-based detection to DRAM and cache coherence attacks (unfortunately, evaluating these is beyond our scope due to space limitations). Compared to naively counting flushes, counting flushes-on-zombies could lower the false-positives in attack detection -- as this can detect episodes of repeat flushes to the same line (common to such attacks)
\ignore{
Note that ZBM does not exacerbate main-memory attacks. Such attacks, that require repeated memory accesses, need eviction of a line from the private caches (with \texttt{clflush}) on each iteration, with or without ZBM. In fact, Zombie can help detect such attacks that need frequent usage of \texttt{clflush}. Compared to naively counting flushes, counting flushes-on-zombies could lower the false-positives -- as this can detect episodes of repeat flushes to the same line that is common to such attacks. Unfortunately, evaluating these attacks is beyond the scope of this paper, due to space limitations.
}
\subsubsection{Tolerating DRAM Timing Side-Channels} 
DRAMA~\cite{DramaRowbuffer} exploits DRAM row-buffer timing as a side-channel. This attack uses the timing difference between row-hit and row-miss to leak the access pattern. To execute at a sufficiently fast rate, the attack requires flush of the cache line in each iteration to ensure that the data is always read from memory. As these attacks incur frequent episodes of Zombie-Miss and then Flush-on-Zombie, we can count such episodes with dedicated hardware counters and detect a potential attack. When the counters cross a threshold, mitigation can be performed by switching to a closed-page policy.

\ignore{performance and power benefits of open-page policy in benign scenario.}

\subsubsection{Tolerating DRAM Row-Hammer Attacks}

Row-Hammer attacks~\cite{kim:isca14} flush two cache lines in each iteration, each corresponding to a different row in the same bank of DRAM. This causes frequent accesses to alternate rows, and a large number of row closures on the same rows within a short time-period. This injects faults in other neighbouring rows (due to faulty isolation in DRAM technology). Adversaries leverage this to inject bit-flips and modify access control bits in sensitive kernel data-structures, engineering illegal privilege escalation. While this attack can be mitigated by increasing the DRAM refresh rate\ignore{ from 64ms to 32ms/16ms}, such a solution incurs performance and power overheads. Instead, as these attacks incur frequent episodes of Zombie-Miss and then Flush-on-Zombie, detection is possible with dedicated hardware counters tracking these episodes. When the count exceeds a threshold, mitigation can be performed by increasing the DRAM refresh rate for a certain time period.

\subsubsection{Tolerating Coherence-Based Attacks}

OS-shared pages can leak information through the coherence status of their cached lines, as per recent attacks~\cite{yao:coherence}. Such attacks exploit the difference in cache access latency, based on whether the line is in Exclusive (\textit{E}) or Shared (\textit{S}) state. The spy first initializes the line to \textit{E}, then allows the victim to execute.  Later, by checking if the line is in \textit{S} (based on access latency), the spy can infer if the victim accessed it.

For this attack, re-initializing the line to \textit{E} is essential on each iteration: achieved with a {\texttt{clflush}} to invalidate the line, and a subsequent load to re-install it in \textit{E} (note: using stores to invalidate copies of lines leveraging coherence is not useful as that would unshare OS-shared pages). Hence, detecting episodes of repeated flush and reload to the same line, using hardware counters for Zombie-misses and Flushes-on-Zombies, can help detect such attacks. When the counts cross a threshold, mitigation can be activated in hardware by servicing  \textit{S} and \textit{E} accesses with similar (worst-case) latency.

\ignore{
\subsection{Tolerating DRAM Timing Attacks}


Recent memory attacks have used cache-line flushes to orchestrate fault-injection or memory-timing attacks. DRAMA~\cite{DramaRowbuffer} exploits DRAM row-buffer timing as a side-channel. This attack uses the timing difference between row-hit and row-miss to leak the access pattern. To execute at a sufficiently fast rate, the attack requires flush of the cache line in each iteration to ensure that the data is read from memory in each iteration. While this attack can be mitigated by switching to a closed-page policy, ideally, we would like to have the performance and power benefits of open-page policy and switch to closed-page policy only under attacks.  Our solution can detect DRAMA attacks, as these attacks incur frequent episodes of Zombie-Miss. We can dedicate a counter to keep track of the number of episodes of Zombie-Miss, and  when the number of such events exceed a threshold, mitigation can be performed by switching to a closed-page policy.

\subsection{Tolerating DRAM Row-Hammer Attacks}

Row-Hammer attacks~\cite{kim:isca14} flush two cache lines in each iteration, each corresponding to a different row in the same bank of DRAM memory (for example, Row-A and Row-B). This causes frequent access to the two rows, and large number of row closures on the same rows of memory within a short period of time. Such an attack injects faults in rows neighboring to the rows being accessed (due to faulty isolation in DRAM technology), and attempts to modify access control bits, to obtain higher privilege illegally. Row-Hammer attacks can be mitigated by reducing the DRAM refresh time from 64ms to 32ms/16ms; however, such a solution incurs performance and power overheads.  Our Zombie-based monitoring can detect Row-Hammer attacks, as these attacks incur frequent episodes of Zombie-Miss. We can dedicate a counter to keep track of the number of Zombie-Miss, and  when this number exceeds a threshold, mitigation can be performed by increasing the DRAM refresh rate for a certain time period.


\ignore{
An adversary can use flushing of shared lines to create alternative attacks to observe victim's access patterns. For example, Flush+Flush~\cite{FlushFlush} attacks use the variation in latency of the flush instruction, depending on whether the line is resident in the cache or not, to infer whether it was accessed by a victim. Such an attack can be mitigated using constant time implementation of clflush instruction. Furthermore, we can also use our insights about monitoring zombie lines to detect and mitigate such attacks. For example, for a Flush+Flush attack to be successful, a flush instruction must cause invalidation of a recently invalidated line (at least a few times). We can use the events of Flush operations to invalid-zombie lines to detect the Flush+Flush attacks and provide mitigation in the hardware or the OS.
}

\ignore{
\subsection{Mitigating Flush-Based Covert Channel Attacks}
}

\subsection{Tolerating Coherence-Based Attacks}
\label{sec:coherence_attacks}

We deem a line to be a zombie only if it gets invalidated due to a Flush, and invalidations due to coherence operations do not create zombie lines. Therefore, read-write sharing of data between different threads of a multi-threaded program (such as ping-pong on a spin lock) does not affect the zombie status of a line. Such read-write shared pages that belong to only one process are not the target of the cross-core Flush+Reload attacks, which try to leak information across two independent processes with only OS-shared pages in common.


Even though such OS-shared pages are used in a read-only mode by user processes, an attacker~\cite{yao:coherence} can use the coherence status of these lines to infer the access patterns in a multi-node system. Such attacks exploit the difference in cache latency, depending on whether the line is in Exclusive (E) or Shared (S) state. By initializing the line to state-E, the spy can infer an access by the victim, as the line would transition the line to S-state on a victim access.

Our solution can detect such coherence-based attacks. For such attacks, initializing the line to state-E is an essential component for each attack attempt, that is achieved by executing a {\texttt{clflush}} to invalidate the line, before a subsequent load to install the line in state-E. With our solution, the flush operation would mark the line as a zombie. Counting subsequent flushes to zombie lines can help detect the attack. It can then be mitigated in hardware by enforcing constant-time accesses independent of coherence state.

}

\section{Related Work}
\label{sec:related}


Our proposal leverages the insight that when a cache line is invalidated on a flush, the tag and the data are still in the cache and can be used for detecting and tolerating flush-based attacks. It is partly inspired by past work on {\em Coherence Decoupling}~\cite{incoherence:asplos04} that used the resident tag and data of the lines invalidated due to coherence, for performing speculative operations (using stale values) while waiting for the coherence response.  In this section, we discuss cache attacks in general and prior solutions to tolerate Flush+Reload attacks.


\ignore{
Note that our goal was to develop a hardware-based solution that not only provides strong security but also (a) has low performance overheads (b) has low storage overheads (c) requires no OS or software support, and (d) can be implemented in a localized manner at the cache controller, without requiring changes to the other parts or constraining the design choices for the memory system.  We use these set of requirements for comparison with CEASER, noting that CEASER fulfills all these requirements. 
}





%

\subsection{Types of Cache Attacks}
Recent works~\cite{CacheAttackSurvey:Yarom,CacheAttackSurvey:CCS} have surveyed both Conflict-Based Attacks~\cite{PrimeProbe,EvictPrefetch,AliasDriven} that generate cache-conflicts to invalidate cachelines and Flush-Based Attacks~\cite{yaromFlushReload,FlushFlush,CoherenceInvalidate,yao:coherence} that use flushes to invalidate cachelines. These have been used to attack algorithms like AES~\cite{CacheGames,AESAccessDriven}, RSA~\cite{RubySP2015,CacheMissFun,CachebleedYarom}, etc. in cryptographic libraries and leak the secret keys.

\subsection{Tolerating Conflict-Based Attacks}
Conflict-based attacks can be tolerated by either preserving the data of the victim  or through randomization. The examples of preservation-based approaches includes PLCache~\cite{RPCache}, CATalyst~\cite{Catalyst}, StealthMem~\cite{StealthMem}, SecDCP~\cite{SECDCP} and  Non-Monopolizable (NoMo) Cache~\cite{NoMo}.  Recent studies, such as Relaxed Inclusion Cache (RIC)~\cite{RIC} and SHARP~\cite{SHARP}, have also targeted the inclusion property of LLC to provide preservation. The examples of randomization-based approaches to detect or mitigate such attacks include ReplayConfusion~\cite{ReplayConfusion}, RPCache\cite{RPCache}, NewCache~\cite{NewCache}, and CEASER~\cite{CEASER}. However, such solutions are ineffective at guarding against flush-based attacks on OS-shared lines as the adversary can evict the  line explicitly without any conflict or inclusion violation.


\subsection{Tolerating Attacks with Software Support}
Prior studies propose rewriting software to avoid storing critical information (such as encryption tables) in memory and instead compute it on-the-fly~\cite{PrimeProbe, SoftwareMitigation}. Unfortunately, such revised implementations tend to be 2x to 4x slower than the original implementation~\cite{SoftwareMitigation}. Our solution avoids the software engineering effort and slowdown of such methods.

Another proposal~\cite{CSD} prefetches \textit{all} the sensitive probe-addresses into the cache (e.g. entire encryption tables) based on the software context (e.g. before running AES), preventing an attacker from selectively observing lines accessed by a victim. However, this cannot scale to prefetch larger shared libraries and requires SW-rewrite to trigger such prefetch.  

\ignore{
\subsection{Tolerating Attacks by Rewriting Software}
Prior studies propose rewriting software to avoid storing critical information (such as encryption tables) in memory and instead compute it on-the-fly~\cite{PrimeProbe, SoftwareMitigation}. Unfortunately, such revised implementations tend to be 2x to 4x slower than the original implementation~\cite{SoftwareMitigation}. Our solution avoids the software engineering effort and slowdown of such methods.
}



\ignore{
Cloak~\cite{tmem:usenix2017} design proposes to rewrite safety-critical software using transactional memory semantics, which means transactions that have concurrent memory accesses to shared location by other processes will cause transaction abort and avoid leakage of timing information to concurrently running applications. This proposal incurs significant software effort, and is suitable only for applications whose working set can fit within the storage of the hardware transactional memory. 
}

\ignore{
Cache attacks can be mitigated by rewriting the security-sensitive applications to not keep critical information (such as encryption tables) in memory and instead use mathematical operations to compute the critical information on-the-fly\cite{PrimeProbe, SoftwareMitigation}. Unfortunately, such revised  software implementation tends to be 2x to 4x slower than the original implementation~\cite{SoftwareMitigation}. Our solution avoids the effort in identifying critical pieces, rewriting the software, and the slowdown of such methods.
}
\subsection{Tolerating Attacks by Restricting {\texttt{clflush}}}

Flush+Reload attacks rely on the use of {\texttt{clflush}} instruction for flushing the OS-shared (read-only) lines. SHARP~\cite{SHARP}  advocates restricting the use of {\texttt{clflush}} in user mode on such read-only lines, and triggering a {\em copy-on-write} through the OS if such an operation is attempted. Unfortunately, such an approach requires changes to the ISA to redefine the {\texttt{clflush}} instruction, is not backwards compatible, and needs modifications to the OS to handle a trap when a clflush is issued to deduplicated pages from a user-space application. Ideally, we want to mitigate the attack in hardware, without any changes to the ISA or to the OS.

\ignore{
Flush+Reload attacks rely on the use of {\texttt{clflush}} instruction for flushing the OS-shared lines between the two different processes, and such sharing of OS-pages typically occurs in read-only mode. SHARP~\cite{SHARP}  advocates restricting the use of {\texttt{clflush}} in user mode, so that OS-shared read-only pages cannot be flushed using {\texttt{clflush}} by a user-mode application.  Unfortunately, such an approach requires changes to the ISA to redefine the {\texttt{clflush}}  instruction, is not backwards compatible, and may cause failure of older user applications. For example, OS typically implements page deduplication by providing the user applications with a read-only copy of the duplicated pages and performs a {\em copy-on-write} when one of the process tries to write to the de-duplicated page. With the restriction placed by SHARP, the user application will not be able to flush a line from any of the de-duplicated page (which could be any page in the user space) unless OS implementation of de-duplication is also updated. Ideally, we want to tolerate the attacks without requiring any ISA/OS changes.
}

\subsection{Tolerating Attacks with OS-based Solutions}
Flush-based attacks can be avoided by disabling sharing of critical pages in OS (as in CATalyst~\cite{Catalyst} or Apparition~\cite{Apparition})  or replicating shared pages on concurrent access by multiple processes\cite{CacheBar}. Unfortunately, such solutions give up the capacity benefits of page sharing and are not preferable.

Prior studies~\cite{PerfCtr1,PerfCtr2,pokerface} have suggested using hardware performance counters or monitoring memory bandwidth~\cite{pokerface} for anomaly detection. Unfortunately, such proposals suffer from false positives and are not universally applicable, as profile information is not always available for all applications.

\ignore{
Flush-based attacks can be avoided by extending the OS to disable sharing of critical pages. CacheBar~\cite{CacheBar} is a scheme that dynamically replicate shared pages if multiple pages are concurrently accessing such pages, thus giving up on the capacity benefits of page sharing even for benign applications.
}
\ignore{
Prior studies~\cite{PerfCtr1,PerfCtr2} have also suggested using hardware performance counters to observe deviation in behavior of critical applications to detect attacks, assuming profile information of the applications (during attack-free scenario) is available. Poker Face Mitigation~\cite{pokerface} exploits the observation that the memory bandwidth consumption increases under Flush+Reload attack, and use this to provide mitigation. Unfortunately, profile information and bus-usage is not always available for all applications limiting the use of such schemes.
}

\subsection{Tolerating Attacks by Line Duplication}

Recent work~\cite{Ruby:MICRO2017} evaluated NewCache~\cite{NewCache} for Flush+ Reload attack by accessing NewCache with Process-ID and line-address. This creates duplicate copies of the line in the cache if two different processes concurrently access the same line, thereby preventing the flush of one process from evicting the line of another.  Unfortunately, such in-cache duplication is incompatible with inter-process communication. Furthermore, NewCache requires storage for mapping tables and the OS to classify applications into protected and unprotected categories, to protect the mapping table from attacks.


DAWG~\cite{DAWG} and MI6~\cite{MI6} allocate a given number of ways and sets respectively per security domain. They protect against Flush+Reload attack by creating a duplicate of the OS-shared line for each domain or disabling page-sharing across domains, thereby preventing the flush of one domain from evicting the line of another. Such a design places restrictions on inter-process communication through shared-memory, in that the communicating processes must be located in the same domain. Moreover, OS support is required for both cache allocation and for assigning processes to security domains.  Ideally, we seek a hardware solution that does not require OS support and does not restrict inter-process communication.

\ignore{

\begin {table}[ht]
  \begin{small} 
  \begin{center}
  \vspace{-0.15in}
      \caption{Storage Overheads (for 16MB LLC). }
        \begin{tabular}{|c|c|} \hline
Scheme              & Storage Overheads \\ \hline \hline
NewCache  (with OS support)    & 2.5 megabytes \\ \hline
NewCache  (without OS support) & 17 megabytes   \\ \hline \hline
ZBM (No OS support required) & 1 bit-per-line (32KB)      \\ \hline
        \end{tabular}
      \label{table:TBR}
       \vspace{-0.15 in}
    \end{center}
  \end{small}
\end{table}
}

\section{Conclusion}
\label{sec:summary}

Recent vulnerabilities exploiting cache side-channels have universally impacted the entire computing industry, underscoring the importance of mitigating them in next-generation hardware.  In this paper, we investigate solutions for tolerating cross-core Flush+Reload attacks. Prior solutions for mitigating flush-based attacks require either rewrite of the application, or OS support, or changes to the ISA.  Ideally, we seek a solution that can efficiently mitigate Flush+Reload attacks without requiring any changes to the software or the OS or the ISA, and while incurring negligible overheads. 

In this paper, we propose a simple hardware mitigation of Flush+Reload cache attacks using {\em Zombie} lines -- lines invalidated by a Flush, but with the tag and data still resident in the cache. Our solution is based on marking zombies on flushes and protecting them till such time they would have been naturally evicted. Our {\em Zombie-Based Mitigation (ZBM)} mitigates the attack in hardware by servicing zombie hits with the same latency as cache misses, thereby avoiding any timing leaks. Moreover, ZBM and its extension ZBMx (for niche applications with frequent flushes) avoid any slowdown for benign applications and incur a storage overhead of only 1-4 bits per cache line. Thus, ZBM is an effective, yet practical mitigation for cross-core Flush+Reload attacks. 


\ignore{
This paper proposes a solution to efficiently tolerate Flush+Reload attack by exploiting the observation that when the line is invalidated on a Flush, the tag and the data of the line are still resident in the cache. We deem such lines as Zombie lines. Our design is based on marking the zombies, protecting them until the line for a time until it would have been naturally evicted from the cache, and provide actions on zombie misses and zombie hits.  Our {\em Zombie-Based Mitigation (ZBM)} mitigates Flush based attacks in hardware by simply servicing zombie hits with a latency similar to cache mises, thereby avoiding any timing leaks. The solution mitigates Flush+Reload attack while incurring a storage overhead of only 1 bit per cache line, and incurring no performance overhead for benign applications.We also discuss how the zombie-based detection can be used to enable alternative mitigation policies in hardware, and for tolerating other attacks.
}



\newpage

\bibliographystyle{ieeetr}
\bibliography{ref}

\end{document}